\documentclass[12pt,a4]{article}

\usepackage{lmodern,sansmath}
\usepackage{textcomp}
\usepackage{microtype}
\usepackage{amsmath, amssymb, amsthm, dsfont, bm}         
\usepackage{amsfonts}         

\usepackage[round]{natbib}
\usepackage[colorlinks=true,linkcolor=red,citecolor=blue]{hyperref}

\usepackage{color}

\usepackage{booktabs}
\usepackage{graphicx}

\usepackage{caption}
\usepackage{subcaption}

\usepackage{rotating}

\newcommand*\dsk{./}

\newcommand*\figpath{\dsk}

\usepackage{latexsym}

\usepackage{url}
\usepackage{xcolor}
\definecolor{newcolor}{rgb}{.8,.349,.1}

\newcommand{\ie}{{\em i.\thinspace{}e. }}

\newcommand{\ds}{\displaystyle}

\newcommand\bet{{g}}

\newcommand\alps{{\frac{\hbar^2}{2m}}}
\newcommand\alpss{{\frac{\hbar}{2m}}}

\newcommand\dertt[1]{ \frac{\partial{ #1}}{\partial t} }
\newcommand\gd{\mbox{${\bf \nabla}^{2}$}}
\newcommand\psib{\overline{\psi}}

\usepackage{color}

\begin{document}



\title{Coupling Navier-Stokes and Gross-Pitaevskii  equations for the numerical  simulation of two-fluid quantum flows}%

\author{Marc Brachet$^1$,  Georges Sadaka$^2$, Zhentong Zhang$^2$,\\
	Victor Kalt$^2$ and Ionut Danaila$^{2, *}$
	\\ \\
	$^1$ENS, Universit\'e PSL, CNRS, \\
	Sorbonne Universit\'e, Universit\'e de Paris, \\
	Laboratoire de Physique de l'\'Ecole Normale Sup\'erieure, \\
	F-75005 Paris, France.
	\\ 
	$^2$Univ Rouen Normandie, CNRS,  \\
	Laboratoire de Math{\'e}matiques Rapha{\"e}l Salem,\\
	  UMR 6085, F-76000 Rouen, France\\
	\\ 
	$^*$ Corresponding author: ionut.danaila@univ-rouen.fr
}

\date{\today}
\maketitle

\begin{abstract}
Numerical methods for solving the Navier-Stokes equations for classical (or normal)  viscous fluids are well established.
This is also the case for the Gross-Pitaevskii equation, governing quantum inviscid flows (or superfluids) in the zero temperature limit. 
In quantum flows, like liquid helium II at intermediate temperatures between zero and 2.17 K, a normal fluid and a superfluid coexist with independent velocity fields. The most advanced existing models for such systems use the Navier-Stokes equations for the normal fluid and 
a simplified description of the superfluid, based on the dynamics of quantized vortex filaments, with ad hoc reconnection rules. 
There was a single attempt (C. Coste, The European Physical Journal B - Condensed Matter and Complex Systems, 1998) to couple Navier-Stokes and Gross-Pitaevskii equations in a global model intended  to describe the compressible two-fluid liquid helium II. 
We present in this contribution a new numerical model to couple a Navier-Stokes incompressible fluid with a Gross-Pitaevskii superfluid. Coupling terms in the global system of equations involve new definitions of the following concepts: the regularized superfluid vorticity and velocity fields, the friction force exerted by quantized vortices to the normal fluid, the covariant gradient operator in the Gross-Pitaevskii based on a slip velocity respecting the dynamics of vortex lines in the normal fluid. A numerical algorithm based on pseudo-spectral Fourier methods is presented for solving the coupled system of equations.
Finally, we numerically test and validate the new numerical system against well-known benchmarks for the evolution in a normal fluid of different types or arrangements of quantized vortices (vortex crystal, vortex dipole and vortex rings). The new coupling model has the advantage to keep the full Gross-Pitaevskii model for the superfluid, and thus describe quantized vortex dynamics without any phenomenological approximation. This opens new possibilities to revisit and enrich existing numerical results for complex quantum fluids, such as quantum turbulent flows.

\end{abstract}


%
%


 
\section{Introduction}\label{sec:intro}

Realistic numerical models for quantum flows, such as liquid helium below the critical (lambda) temperature $T_\lambda =2.17K$, have to accommodate with the celebrated  two-fluid model \citep{Tisza_1938,Landau1941} stating that two fluids with independent velocity fields coexist in the system: a {\em normal} viscous fluid and an inviscid {\em superfluid}. If each component is taken separately, governing equations and numerically models are universally accepted: the Navier-Stokes (NS) equations for the normal fluid (the only one present in helium if $T>T_\lambda$) and the Gross-Pitaevskii (GP) equation for the superfluid (dominant for $T<0.3 K$).  For intermediate temperatures, both components are present, with different physics. In the superfluid, quantized vortices are  nucleated with fixed (quantized) circulation  and fixed core diameter (of the atomic size). Complex interactions between quantized vortices tangled in space can lead to Quantum Turbulence (QT). In the normal fluid, vortices or eddies can develop covering characteristic scales from the Kolmogorov viscous scale up to the container size and eventually generate turbulence. Note that the GP equation is known to well describe low-$T$ co-flow QT \citep{Nore97a,Nore97b,Clark17,KOBAYASHI2021} and can be extended to describe helium equation of state and dispersion law at zero temperature \citep{Berloff_2014}.

Existing numerical models for quantum two-fluid flows either focus on a single component (using NS or GP models) or simplify the physics of one component in the two-fluid setting. The Hall--Vinen--Bekharevich--Khalatnikov (HVBK)  model \citep{Bekarevich_Kalat_1961,HV1_1956,HV2_1956}  describes the normal fluid motion by the NS model, while the superfluid motion is simplified to an Euler-like equation.  
The two fluids are coupled through a friction force that takes into account the influence of quantized vortices through a coarse-grained averaged superfluid vorticity. 
The average is considered  over an ensemble of parallel (polarized) vortex filaments to find an equivalent solid-body vorticity for a dense vortex bundle of line density.
 The individual dynamics of quantized vortices thus disappears in the  HVBK model. 
 
 A different trade-off is made in Line Vortex Navier-Stokes (LV-NS)  models.  Quantized vortices are described as geometrical lines, with infinite velocity and singular vorticity at the centreline. Vortex lines are moved following the Biot-Savart law  and phenomenological models for vortex reconnection are applied. 
  Mutual friction is described, in a more mesoscopic way, by assessing the interaction between the normal NS fluid and the vortex line dynamics  \citep{Tsubota_2010,kivotides2000triple,Giorgio2020}.  These  phenomenological LV-NS models typically contain two ingredients. First, on the vortex lines, a {\bf slip velocity} is added to the Biot-Savart line velocity. This slip velocity is obtained as a function of the counterflow (the difference between the normal fluid velocity on the line and the line velocity) by a standard argument based on the balance of friction and Magnus forces. Second, a spatially smoothed {\bf friction force}, opposite to the friction force acting on the line, is added as a source term in the NS equations.
These vortex line models are limited with respect to both vortex reconnection (that is added to the model in an ad hoc manner) and to vortex nucleation (that is simply non-existent in these models). 

The present contribution is a first attempt, to the best of our knowledge, to directly couple incompressible NS and GP models and thus numerically simulate, without any simplification, the dynamics of a two-fluid quantum flow.  The model is inspired by existing LV-NS models  from which we extract the main physical ingredients of the mutual friction  produced by the interaction of the normal fluid and  superfluid vortices. The novelty is that we transpose this mutual friction coupling into the framework of the GP model that has the advantage to describe vortex nucleation and vortex interactions without any phenomenological assumptions \citep{Koplik93,FPR92}. We develop consistent expressions for the coupling with new  interaction terms and we prove  numerically that they are compatible with known phenomenological mutual friction laws. We first derive a {\bf regularized line velocity field} that is smooth and reduces to the value of the line velocity, when evaluated on the vortex line. Using this regularized velocity field we build  a slip velocity field that is used to couple GP and NS equations. As a main consequence of this study, coupling of NS and GP numerical codes becomes possible with this new GP-NS model.

We should mention that, in the different context of Landau's original {\bf compressible} two-fluid model  \citep{Balibar2017}  describing second sound and containing neither vortices nor mutual friction, Coste  \citep{Coste1998} studied ways to couple NS and GP equations. Nevertheless, an outcome of Coste's model was to introduce a simple coupling law of the local counterflow vector to the GP equation. In the following we will use a coupling that is closely related, but different, to the one pioneered by Coste.

The paper is organized as follows. The theoretical background is given in Section  \ref{sec:theorybg}. After first defining the uncoupled GPE and NSE equations in Section \ref{sec:theory_uncoipled} the coupling terms are derived in Section \ref{sec:ModelDef}. The numerical implementation is described in Section \ref{sec:Impl}. Our results are contained in Section \ref{sec:Numr} and, finally, Section \ref{sec:concl} is our conclusion.

\section{Theoretical background} \label{sec:theorybg}

\subsection{The uncoupled GP and NS equations} \label{sec:theory_uncoipled}


The GP equation is a partial differential equation 
describing the dynamics of a dilute
superfluid Bose-Einstein condensate
at zero-temperature.
It applies to a complex field $\psi$, where $|\psi|^2$ is the number of particles per unit volume, and reads
\begin{equation} 
	i\hbar\dertt{\psi}  =- \alps \gd \psi + \bet|\psi|^2\psi,
	\label{Eq:GPE}
\end{equation}
where $m$ is
the mass of the condensed particles, $\hbar$ the reduced Planck constant, and $g$ the interaction constant with $g=4 \pi  \tilde{a} \hbar^2/m$ and
$\tilde{a}$  the $s$-wave scattering length. 

Equation~(\ref{Eq:GPE}) can be mapped into hydrodynamic
equations for a compressible fluid by the Madelung transformation 
\begin{equation}
	\psi({\bf x},t)=\sqrt{\frac{\rho({\bf x},t)}{m}}\exp{\left(i
		\frac{m}{\hbar}\phi({\bf x},t)\right)},\label{Eq:defMadelung}
\end{equation}
where $\rho({\bf x},t)$ is  the  mass density of the fluid, $\phi({\bf x},t)$  the velocity potential 
associated to the fluid velocity 
${\bf v}=\frac{\hbar}{m}{\bf \nabla} \phi$.
This transformation is singular on
the zeros of $\psi$. As two conditions are required 
(both real and imaginary parts of $\psi$ must vanish), these
singularities generically take the form of points in 2D and 
lines in 3D. The Onsager-Feynman quantum of velocity circulation around
vortex lines with $\psi=0$ is  $\Gamma=h/m$. Thus, due to the multivalued nature of the velocity potential in the presence of vortex lines, 
the superflow is not irrotational. It can be proved \citep{Clark17}, using the Madelung transformation, that
the vorticity  ${\bf \omega}=\nabla \times {\bf v}$ is given by
\begin{equation}
	{\bf \omega} ({\bf
		r}) = \frac{h}{m} \int d s \frac{d {\bf r}_0}{d s} \delta({\bf r} -
	{\bf r}_0(s)), \label{eq:vortdelta}
\end{equation}
where ${\bf r}_0(s)$ denotes the position of the vortex line, $\delta$ is the Dirac delta function and $s$ the arclength. 
Thus, the vorticity in a quantum flow is a
distribution concentrated along the $\psi=0$ topological line defects where
${\bf v}$ is ill-behaved (with a $1/r$ divergence).

Linearizing the GP equation around a constant state
$\psi=\Psi_0$ yields the Bogoliubov dispersion relation
for density plane waves ($\rho_0 e ^{i ({\bf k}\cdot {\bf x} - \omega t)}$, with ${\bf k}$ the wave number vector):
\begin{equation} 
	\omega_B(k)=\sqrt{\frac{g {\bf k}^2 |\Psi_0|^2}{m}+\frac{\hbar^2 {\bf k}^4}{4
			m^2}}.
	\label{eq:Bog}
\end{equation} 
The sound velocity is thus given by 
\begin{equation}
c=\sqrt{g|\Psi_{0}|^2/m}. \label{Eq:defc}
\end{equation}
Dispersive effects take place for length scales smaller than the 
coherence length, defined by
\begin{equation}
	\xi=\frac{\hbar}{\sqrt{2gm|\Psi_{0}|^2}}. \label{Eq:defxi}
\end{equation}
Note that $\xi$ is proportional to the radius of the vortex core
\citep{Nore97a,Nore97b}.

The GP equation conserves
the total energy $E$, the total mass $\mathcal{M}$, and
the momentum ${\bf P}$, which are defined in a volume $V$ as
\begin{eqnarray}
	E&=&\int_{V}  \left( \alps |\nabla \psi |^2
	+\frac{g}{2}|\psi|^4 \right)\,d^3x , \label{Eq:defH}\\
	\mathcal{M}&=&m\int_V  |\psi|^2 \,d^3x ,\label{Eq:defN}\\
	{\bf P}&=&\int_V  \frac{i\hbar}{2}\left(
	\psi {\bf \nabla}\psib - \psib {\bf
		\nabla}\psi\right)\,d^3x\label{Eq:defP},
\end{eqnarray}
where the overline denotes the complex conjugate.

To describe  the dynamics of  a viscous incompressible flow of velocity vector field $\bf{v}$
we use the Navier-Stokes equations
\begin{align} \label{eq:NSE-m}
	\partial_t {\bf v}+({\bf v}\cdot \nabla){\bf v}=& \ds -\frac{1}{\rho}\nabla p + \nu \nabla^2  {\bf v},\\ 
	\nabla \cdot {\bf v}=& 0, \label{eq:NSE}
\end{align}
where $\rho$ is the constant flow density, $\nu$ the kinematic viscosity and $p$ denotes the pressure field that enforces incompressibility (\ie zero divergence velocity field).

The NS  equations \eqref{eq:NSE-m}-\eqref{eq:NSE} conserve the total mass and the total momentum and, only for inviscid flows (with $\nu=0$) the energy is also conserved:
\begin{eqnarray}
	E&=&\rho \int_{V} \,\frac{{\bf v}^2}{2} \,d^3 x , \label{Eq:defHNS}\\
	\mathcal{M}&=&\rho \int_V  \,d^3x ,\label{Eq:defMNS}\\
	{\bf P}&=& \rho \int_V  {\bf v}\, d^3x\label{Eq:defPNS}.
\end{eqnarray}

\pagebreak

\subsection{Building up the model}
\label{sec:ModelDef}
Our reasoning of model building is the following.
In a nutshell, standard phenomenological Line Vortex Navier Stokes (LV-NS) models, such as those developed by \cite{Tsubota_2010,kivotides2000triple,Giorgio2020}, are based on the argument of cancellation of the sum of mutual friction force  and Magnus force acting on the vortex line. The former is 
caused by the difference between the normal fluid velocity and vortex line velocity, while the latter is caused by the slip velocity, \ie the  difference between the vortex line velocity and superfluid velocity. This cancellation yields a phenomenological expression for the slip velocity of the vortex line that is added to the Biot-Savart expression for the equation of motion of the lines. The volume friction force that is added as a source term in the NS equation is then obtained by spatially smoothing the friction on the vortex line.

To apply the same logic to a Gross-Pitaevskii-Navier-Stokes (GP-NS) model, three separate ingredients are needed. The first one is the equivalent of the Biot-Savart velocity of vortex lines: we need a (smooth) field $ {\bf v}^{reg}_s$ (obtained from the GP wave function $\psi$) that, when evaluated on  vortex lines, will give the line velocities  induced by the GP dynamics (in the absence of friction). Second, we need to generalize (using a volume force version) the expression of mutual friction and  Magnus force cancellation. This computation will yield two results: (i) a line slip velocity field ${\bf v}_{\rm slip}$ which reduces on the vortex line to the standard expression used in LV-NS models and (ii) a friction force field $\mathbf{F}_{SN}$ that will be added to the right-hand side of the NS equation. Finally, as a third ingredient, we need an expression for the coupling term in the GP equation that will produce the correct slip velocity ${\bf v}_{\rm slip}$ of vortex lines.  
This coupling term is closely related, but different, to the one pioneered by \cite{Coste1998}.

\subsubsection{The regularized superfluid  velocity field}

The superfluid velocity ${\bf v}_s$ can be simply defined by using the superfluid density 
$\rho_s = |\psi|^2$ and the relation ${\bf P}= \rho_s {\bf v}_s$ for the superfluid momentum, with ${\bf P}$ defined in Eq. \eqref{Eq:defP}.
However, with line vortices present in the GP model, the associated ${\bf v}_s$ can also be estimated by using the Biot-Savart expression stemming from Eq. \eqref{eq:vortdelta}. Since ${\bf v}_s$ has singularities on the vortex line,  we have to introduce a regularized velocity ${\bf v}^{reg}_s$ that is finite on vortex lines and yields the correct velocity circulation at large distances from vortex lines. To wit, we use the following  Gaussian smoothing 
of the physical space field
\begin{align}
	{\bf v}^\epsilon_s({\bf r})&=\frac{i\hbar}{2m}\frac {\psi {\bf \nabla}\psib - \psib {\bf \nabla}\psi}{ \psib \psi+\epsilon^2 \overline{\rho}_s},\nonumber \\
	{\bf v}^{reg}_s&=(1+\epsilon^2)\, {\mathcal F}^{-1}\left(e^{-\frac{k^2}{k_{\rm reg}^2}}{\mathcal F} ({\bf v}^\epsilon_s)\right),\label{eq:vreg}
\end{align}
where ${\mathcal F}$ denotes the Fourier transform and $\overline{\rho}_s=<|\psi|^2>$ is the spatially averaged superfluid density.
The smoothing wave-number parameter  $k_{\rm reg}$ is analogous to the smoothing distance used as a parameter in LV-NS models to obtain the volume force added to NS equations. Parameter $\epsilon$ is used to avoid velocity divergence on the vortex line (where $\psib \psi=0$) and
has to be large enough to correctly resolve vortex lines. 
In practice (see Section \ref{sec:Impl}), we set $\epsilon^2=0.1$ and $k_{\rm reg}=1/\xi$.

The regularised velocity field allows one to define a smoothed vorticity, as the curl of the regularized velocity: 
\begin{equation} 
	{\bf \Omega}=\nabla \times \mathbf{v}^{reg}_s.
	\label{eq-Omega}
\end{equation}
For a straight vortex line, the effect of this Gaussian smoothing on the maximum value of smoothed vorticity,
can be estimated to be $F^{-1}$,  given by the  integral 
\begin{equation} 
	F^{-1}=\frac{\hbar}{2m} \frac{1}{\pi}\left[\int_{-\infty}^\infty  e^{-\frac{k^2}{k_{\rm reg}^2}}  \, dk\right]^2 = \frac{\hbar}{2m} \, k_{\rm reg}^2.
\end{equation}
We finally define the 'normalized' vorticity field
\begin{equation} 
	{\bf \hat \Omega}=F {\bf \Omega}=F \nabla \times \mathbf{v}^{reg}_s, \label{eq:omegahat}
\end{equation}
which has a norm that is maximum and close to $1$ on the vortex line and  much smaller than $1$ away from the vortex line.

\subsubsection{Determination of the slip velocity field and volume friction force}

\begin{figure}[!h]
	\begin{center}
		\centering		
						\includegraphics[width=0.75\textwidth]{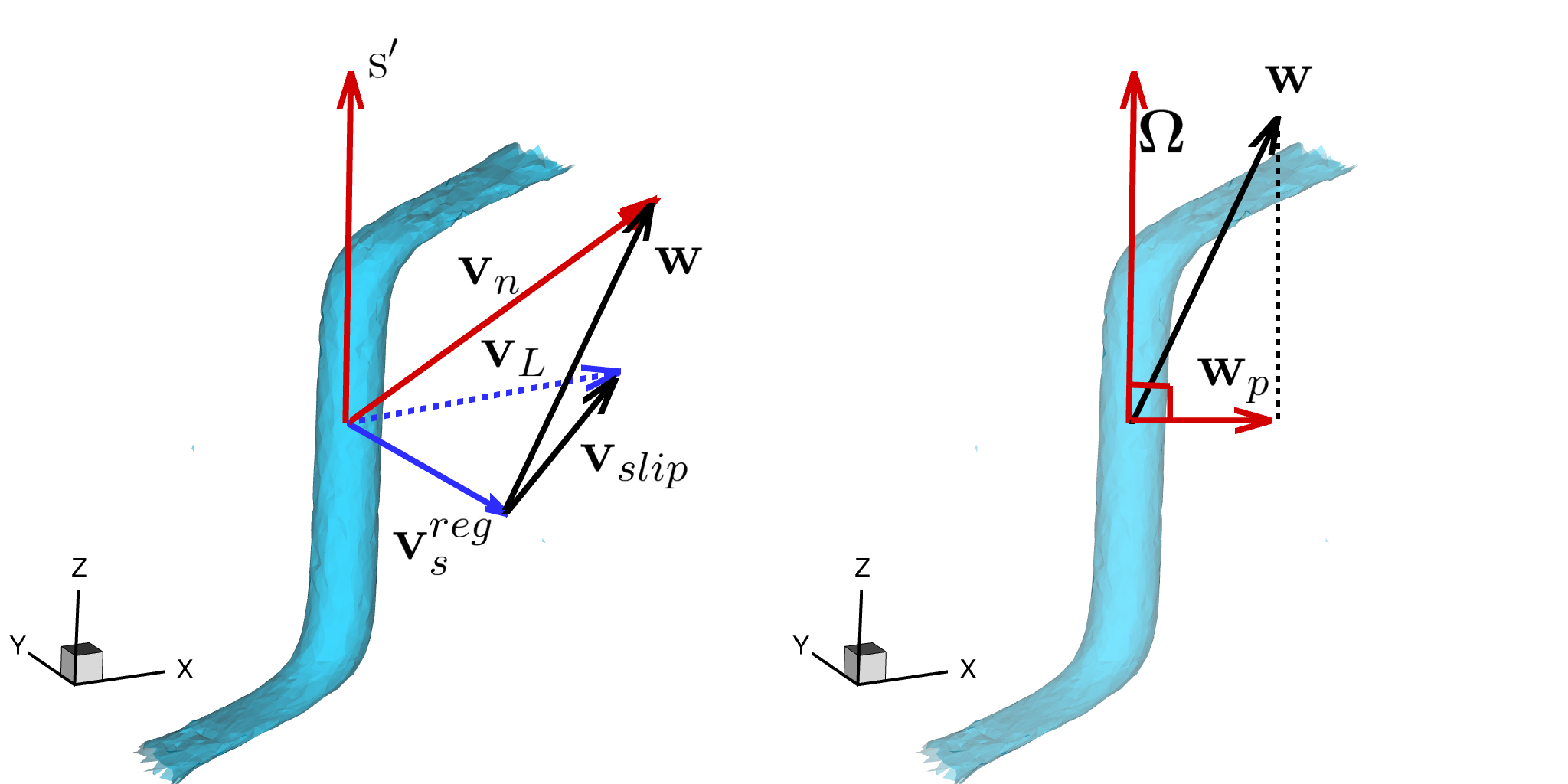}			
		\caption{Sketch of velocities acting on a vortex line. }
		\label{fig:sketch}
	\end{center}
\end{figure}

The Magnus force density caused by ${\bf v}_{\rm slip}$ can be estimated starting from the momentum conservation equation \citep{Sonin1997}:
\begin{equation}
	\mathbf{F}_{MD} = \rho_s \; {\bf v}_{\rm slip}    \times  (\nabla \times \mathbf{v}^{reg}_s).\label{eq:mag}
\end{equation}
This force density must be opposite to the force density acting on the NS fluid, thus
\begin{equation}
	\mathbf{F}_{MD} = -\mathbf{F}_{SN}.\label{eq:fsnmag}
\end{equation}
For  $\mathbf{F}_{SN}$ we start from the simple phenomenological expression considering a force with longitudinal and transversal components 
\begin{equation}
	\mathbf{F}_{SN} \sim \rho_n \left[\beta   s' \times ( s' \times ({\bf v}_{n}-{\bf v}_{L}))+ \beta' s' \times ({\bf v}_{n}-{\bf v}_{L})\right],
	\label{eq:pheno-exp-01_3}
\end{equation}
where $\rho_n$ and ${\bf v}_n$ are the density and velocity of the normal fluid,  ${\bf v}_L$ the velocity of the vortex line, $s'$ the unit tangent to the line (see Fig. \ref{fig:sketch}), and  $\beta$, $\beta'$ two phenomenological coefficients.

Using the fact that {\bf on vortex lines} the vector ${\bf \hat \Omega}=F {\bf \Omega}$ is of norm $1$ and directed along the line, we postulate the following formula for the volume force, equivalent to Eq. \eqref{eq:pheno-exp-01_3}:
\begin{equation}
	\mathbf{F}_{SN}= \rho_n   \left[ B_\star (\nabla \times \mathbf{v}^{reg}_s) \times  ( F(\nabla \times \mathbf{v}^{reg}_s) \times ( {\bf v}_{n}-{\bf v}_{L}))
	+ B'_\star (\nabla \times \mathbf{v}^{reg}_s) \times ({\bf v}_{n}-{\bf v}_{L})\right], \label{eq:FSN}
\end{equation}
with $B_\star$ and $B'_\star$ the new phenomenological constants. 
After replacing in  \eqref{eq:fsnmag} the expressions  \eqref{eq:mag} and \eqref{eq:FSN},  we need to solve
\begin{align}
	0= - &\rho_s (\nabla \times \mathbf{v}^{reg}_s) \times  {\bf v}_{\rm slip} \nonumber \\
	+&\rho_n  \left[ B_\star (\nabla \times \mathbf{v}^{reg}_s) \times  ( F(\nabla \times \mathbf{v}^{reg}_s) \times ( {\bf v}_{n}-{\bf v}_{L}))
	+ B'_\star (\nabla \times \mathbf{v}^{reg}_s) \times ({\bf v}_{n}-{\bf v}_{L})\right].
	\label{eq-to-solve-1}
\end{align}
Because of the coupling induced slip velocity, the line velocity ${\bf v}_{L}$ is given by (see Fig. \ref{fig:sketch})
\begin{equation}
	{\bf v}_{L}=\mathbf{v}^{reg}_s+ {\bf v}_{\rm slip}.
\end{equation}
Therefore \eqref{eq-to-solve-1} becomes
\begin{align}\nonumber
	0=&{-}\rho_s \; (\nabla \times \mathbf{v}^{reg}_s) \times  {\bf v}_{\rm slip}\\ \nonumber
	+& \rho_n   \left[ B_\star (\nabla \times \mathbf{v}^{reg}_s) \times  ( F(\nabla \times \mathbf{v}^{reg}_s) \times ( {\bf v}_{n}-\mathbf{v}^{reg}_s - {\bf v}_{\rm slip}))\right.\nonumber \\
	+& \left. B'_\star (\nabla \times \mathbf{v}^{reg}_s) \times ({\bf v}_{n}-\mathbf{v}^{reg}_s - {\bf v}_{\rm slip})\right]. \label{eq:vslip33}
\end{align}
A general remark on the equation to solve for $ {\bf v}_{\rm slip}$ is that it involves two vectors fields that are obtained from the normal and superfluid components: the counterflow (see Fig. \ref{fig:sketch})
\begin{equation}
	{\bf w}={\bf v}_{n}-\mathbf{v}^{reg}_s,
\end{equation}
and the (regularized) superfluid vorticity ${\bf \Omega}$ defined in \eqref{eq-Omega}. Recall that ${\bf \Omega}$ is aligned, on  vortex lines, to the vector $s'$ tangent  to the line. 
Supposing that ${\bf w}$ is not aligned with ${\bf \Omega}$, we define the component of ${\bf w}$ perpendicular to the vortex line (see Fig. \ref{fig:sketch}) 
\begin{equation}
	{\bf w}_p={\bf w}-\frac{{\bf w}\cdot{\bf \Omega}}{{\bf \Omega}\cdot {\bf \Omega}}{\bf \Omega}.
\end{equation} 
We note that a natural vector basis is  then $({\bf \Omega},{\bf w}_p,{\bf \Omega}\times{\bf w})$.

Using these variables, Eq. \eqref{eq:vslip33} becomes
\begin{equation}
	0={-}\rho_s \; {\bf \Omega}\times  {\bf v}_{\rm slip} + \rho_n   \left[B_\star {\bf \Omega}\times  ( F{\bf \Omega}\times ( {\bf w} - {\bf v}_{\rm slip}))
	+ B'_\star {\bf \Omega}\times ({\bf w} - {\bf v}_{\rm slip})\right], \label{eq:vslip44}
\end{equation}
which is the same as
\begin{equation}
	0={-}\rho_s \; {\bf \Omega}\times  {\bf v}_{\rm slip} + \rho_n   \left[ B_\star {\bf \Omega}\times  ( F{\bf \Omega}\times ( {\bf w}_p - {\bf v}_{\rm slip}))
	+  B'_\star {\bf \Omega}\times ({\bf w}_p - {\bf v}_{\rm slip})\right].\label{eq:vslip5}
\end{equation}

After some elementary algebraic manipulations (see details in \ref{sec:app_formulae}), we obtain closed expressions 
for the mutual friction, the Magnus force and the slip velocity ${\bf v}_{\rm slip}$. The latter is presented in the following convenient form  
\begin{equation}
	{\bf v}_{\rm slip}=U_\star {\bf w}_p +V_\star {\bf \hat \Omega}\times{\bf w},
	\label{eq-vslip-final}
\end{equation}
with ${\bf \hat \Omega}$ defined in  \eqref{eq:omegahat} 	and
\begin{eqnarray}
		\label{eq-vslip-ustar}
	U_\star &=&\frac{\rho_n \left(B_\star^2 |{\bf \hat\Omega}|^2 \rho_n {+}B'_\star \left(\rho_s{+}
		\rho_n B'_\star \right)\right)}{B_\star^2   |{\bf \hat\Omega}|^2 \rho_n^2+\left(\rho_s{+}
		\rho_n B'_\star \right)^2},\\ 
	V_\star &=&\frac{B_\star \rho_n \rho_s}{B_\star^2 
		 |{\bf \hat\Omega}|^2 \rho_n^2+\left(\rho_s{+} \rho_n B'_\star \right)^2}.
	\label{eq-vslip-vstar}
\end{eqnarray}
Note that the dimensions of the fields in the above expressions are (with $L$, $T$ and $M$ denoting units of length, time and mass): $[{\bf v}_{\rm slip}]=L T^{-1}$; $[{\bf \Omega}]=[\nabla \times \mathbf{v}^{reg}_s]=T^{-1}$; $[F]=T$, thus $[{\bf \hat \Omega}]=1$ and $[\rho]=M L^{-3}$.
In Eq. \eqref{eq:mag}, $\mathbf{F}_{MD}$ is a force per volume: $[\mathbf{F}_{MD}]=M L^{-2}T^{-2}$.
The same dimension is obtained  in Eq. \eqref{eq:FSN} for $\mathbf{F}_{SN}$ because of \eqref{eq:fsnmag}: $[\mathbf{F}_{SN}]=M L^{-2}T^{-2}$.
By inspection, we conclude that the following coefficients are dimensionless: $B_\star$,  $B'_\star$,  $U_\star$, $V_\star$.

To summarize our results, ${\bf v}_{\rm slip}$ is obtained from \eqref{eq-vslip-final}-\eqref{eq-vslip-ustar}-\eqref{eq-vslip-vstar} and the final expression of the friction force results from  \eqref{eq:fsnmag}, \eqref{eq:mag} and \eqref{eq:FSN} as
\begin{equation}
	\mathbf{F}_{SN}=\rho_s \;   {\bf \Omega}\times(U_\star {\bf w}_p +V_\star {\bf \hat \Omega}\times{\bf w}).\label{eq:FSNfinal}
\end{equation}

\subsubsection{Definition of coupling terms in the GP equation}

Expression \eqref{eq:FSNfinal} gives the smooth friction force field to be added to the right-hand side of the NS momentum equation \eqref{eq:NSE-m}. It is apparent by inspection that the $U_\star$ term corresponds to a force normal  to the counterflow ${\bf w}$ and to a slip velocity parallel to ${\bf w}$, while the $V_\star$ term corresponds to a force parallel to ${\bf w}$ and a slip velocity perpendicular to ${\bf w}$. Therefore, on physical grounds, we expect the $V_\star$ term to remove energy from the GP dynamics (and transfer it to the NS flow) while the $U_\star$ term is expected just to change the longitudinal speed of a vortex. This point will be important in our definition of the GP coupling term.

We still need to find a way to implement the slip velocity \eqref{eq-vslip-final} into the GP equation \eqref{Eq:GPE} in a way that will make the vortex lines move with an additional velocity ${\bf v}_{\rm slip}$.
For this purpose, we consider the vortex solution of the stationary GP equation \citep{Nore1997}.  
In 2D polar coordinates $(x=r \cos(\theta), y=r \sin(\theta))$ this solution representing a positive or negative vortex placed at the origin is
\begin{equation}
	\psi_v=R(r) e^{\pm i \theta},
\end{equation}
and satisfies
\begin{equation} 
0=- \alps \gd \psi_v + \bet|\psi_v|^2\psi_v.
	\label{Eq:GPES}
\end{equation}

The time-evolution of this vortex advected by a constant vector field ${\bf U}^{adv}$ 
is described by the partial differential equation
\begin{equation}
\partial_t \psi + {\bf U}^{adv}\cdot \nabla \psi  = i\left( \alpss \gd \psi - \frac{g}{\hbar}|\psi|^2\psi\right),  \label{eq:real}
\end{equation}
with solution
\begin{equation}
	\psi({\bf r},t)=\psi_v({\bf r}-t {\bf U}^{adv}).
\end{equation}
Consider now the advection velocity 
\begin{equation}
	{\bf U}^{adv}_\perp= \pm {\bf \hat e}_z \times {\bf U}^{adv},
	\label{eq-Uadvp}
\end{equation}
where ${\bf \hat e}_z$ denotes  the unit vector in the $z$ direction, and the imaginary-time dynamics
\begin{equation}
	\partial_t \psi - i {\bf U}^{adv}_\perp\cdot \nabla \psi  = i\left( \alpss \gd \psi - \frac{g}{\hbar}|\psi|^2\psi\right). \label{eq:imag}
\end{equation}
Setting ${\bf U}^{adv}=(\cos(\theta^{adv}),\sin(\theta^{adv}))$
and using space and time Taylor series expansions  in both positive and negative vortex cases, the vortex position $(\delta x, \delta y)$ for short times $\delta t$ is given by the solution of the equation
\begin{equation}
	( \delta x + i \delta y - e^{i\theta^{adv}} \delta t ) \frac{dR}{dr}(0) = 0,
\end{equation}  
showing that the position of the vortex is indeed moving with velocity ${\bf U}^{adv}$.

Thus, there are two different ways to move vortex lines in the GP framework with real advection velocity ${\bf U}^{adv}$,  by adding  a  term which is either real \eqref{eq:real} or imaginary \eqref{eq:imag}.

The first approach \eqref{eq:real} corresponds to that suggested by \cite{Coste1998} and is best suited for non-dissipative processes of the GP type. \cite{Coste1998} coupled a vector $\bf{v}$ field to GP dynamics \eqref{Eq:GPE}
through the following substitutions
\begin{equation}
	\nabla \to \nabla + \frac{i}{2 \alpha} \bf{v},
\end{equation}
where we used to short-hand notation $\alpha=\frac{\hbar}{2m}$. The new gradient is similar to the covariant gradient $\nabla + i {\bf A}$ used in magnetic Ginzburg-Landau models, with ${\bf A}$ the electro-magnetic potential vector field \citep{sandier-serfaty}. We notice that 
\begin{equation}
	\alpha {\nabla^2} \to \alpha \nabla^2 + i {\bf v}\cdot \nabla + \frac{i}{2}(\nabla \cdot {\bf v}) - \frac{{\bf v}^2}{4 \alpha},  \label{eq:Coste}
\end{equation}
where the divergence term in the right-hand side  of \eqref{eq:Coste} enforces mass conservation in the modified GP equation,
which becomes
\begin{equation}
	-i\partial_t{\psi}=   \alpha \nabla^2 {\psi}+ i {\bf v}\cdot \nabla {\psi}+ \frac{i}{2}(\nabla \cdot  {\bf v}){\psi}  - \frac{{\bf v}^2}{4 \alpha} {\psi} - \frac{g}{\hbar} |\psi|^2.
	 \label{eq:CosteS}
\end{equation}

Note that, for constant $\bf{v}$, this is a simple Galilean boost with speed  $\bf{v}$.
Indeed, recall that the Galilean invariance of the GP equation explicitly reads:
\begin{equation}
	\psi({\bf x},t) \rightarrow  \psi({\bf x} - {\bf U}^{adv} t,t)
	\exp\left(i \left(\frac{{\bf U}^{adv}}{2 \alpha} \cdot
	{\bf x}-\frac{({\bf U}^{adv})^{2}}{4 \alpha} t\right)\right),
\end{equation}
where ${\bf U}^{adv}$ is the constant velocity of the boost.
This transformation maps any solution $\psi({\bf x},t)$
of the GP equation  into another solution
with associated velocity and density fields that
are Galilean transforms of those associated to $\psi$.
Thus, with $\psi_v ({\bf x})$ denoting as before a time-stationary vortex line solution of the GP equation,  the initial data
$\psi_v ({\bf x}) \exp(i \frac{{\bf U}^{adv}}{2 \alpha} \cdot {\bf x})$
corresponds to a 
vortex translating with (uniform) velocity ${\bf U}^{adv}$.

The second  approach \eqref{eq:imag} is new and related to the damped Schr{\"o}dinger/Gross-Pitaevskii equation \citep{BEC-review-2013-Bao-KRM}, introducing a
dissipative dynamics of the Ginzburg-Landau type.
\cite{Nore97a} prepared an initial data for the GP equation consisting of an array of vortex lines moving at short times with given large-scale velocity field ${\bf U}^{adv}$ by finding a stationary solution of
the Advective Real Ginzburg-Landau Equation (ARGLE):
\begin{equation}
	\partial_t{\psi}=  \alpha \nabla^2 \psi + (\overline{\rho}_s-\frac{g}{\hbar} |\psi|^2)\psi
	-i {\bf U}^{adv} \cdot \nabla
	\psi -\frac{({\bf U}^{adv})^2}{4 \alpha} \psi. \label{eq:ARGL}
\end{equation}
A solution to \eqref{eq:ARGL} corresponds to  a minimum of the associated (modified) GP energy functional:
\begin{equation}
	\label{energie2}
	{\cal{E}}_{ARGLE} [ \psi, \bar \psi]
	=
	\int \left( \alpha \left|\nabla \psi -i \frac{{\bf U}^{adv}}{2 \alpha} \psi\right|^2
	+ \left(\frac{g}{2}|\psi|^4-|\psi|^2\right) \right) \,d^3 {x} .
\end{equation}



We note that the advection term in the ARGLE Eq. \eqref{eq:ARGL}  has opposite sign to the advection term in \eqref{eq:imag}. This means, 
heuristically, that in an ARGLE-converged stationary state with vortices,  the motion that would be created by all vortices is equally balanced by the ARGLE advection term.



Based on these mathematical-physical observations, we conclude that is necessary to split the slip velocity $v_{slip}$, defined in Eq. \eqref{eq-vslip-final}, into  $v_{slip}^\parallel= U_\star {\bf w}_p $ and  $v_{slip}^\perp =V_\star {\bf \hat \Omega}\times{\bf w}$.
For  the coupling with GP equation we use the approach \eqref{eq:real} with ${\bf U}^{adv}=v_{slip}^\parallel$ and \eqref{eq:imag} with ${\bf U}^{adv}_\perp$ given by 
\eqref{eq-Uadvp}. Note that  in our case   ${\bf U}^{adv}_\perp = \pm (\pm \frac{\bf \hat \Omega}{| {\bf \hat \Omega}|}) \times (V_\star {\bf \hat \Omega}\times{\bf w}) = V_\star  | {\bf \hat \Omega}| {\bf w}_p$.

Finally, the expression that has to be used for ${\bf v}$ in the modified GP equation \eqref{eq:CosteS} is
\begin{equation}
	{\bf v}_{\rm slip}^{cpl}=(U_\star + i  V_\star |{\bf \hat \Omega}| ){\bf w}_p.\label{eq:vslipalt}
\end{equation}
Implemented in this way, the coupling corresponding to the perpendicular speed dissipates energy, as it should be,  because of the work of the friction force.

\subsection{Numerical coupling algorithm}\label{sec:Impl}

We start by solving the modified Navier-Stokes equations written in the form:
\begin{eqnarray}
	\partial_t {{\bf v}_n}+({{\bf v}_n}\cdot \nabla){{\bf v}_n}&=&-\frac{1}{\rho_n}\nabla p \nonumber + \nu_n \nabla^2  {{\bf v}_n}+ \frac{1}{\rho_n}{\bf F}_{SN},\\ 
	\nabla \cdot {{\bf v}_n}&=&0, \label{eq:NSEfullNum}
\end{eqnarray}
where 
\begin{equation}
	\mathbf{F}_{SN}=\rho_s \;   (\nabla \times \mathbf{v}^{reg}_s)\times(U_\star {\bf w}_p +V_\star {\bf \hat \Omega}\times{\bf w}),
\end{equation}
with
${\bf w}={\bf v}_{n}-\mathbf{v}^{reg}_s$, ${\bf w}_p={\bf w}-\frac{{\bf w}\cdot{\bf \hat \Omega}}{|{\bf \hat\Omega}|^2}{\bf \hat\Omega}$ and
$U_\star$ and $V_\star$ given by Eqs. \eqref{eq-vslip-ustar} and \eqref{eq-vslip-vstar}, respectively.
%
%
Fields $\mathbf{v}^{reg}_s$  and ${\bf \hat\Omega}$ given by Eq.  \eqref{eq:vreg} and \eqref{eq:omegahat}, respectively, realize the coupling with the modified GP equation  \eqref{eq:CosteS} in which ${\bf v} = {\bf v}_{\rm slip}^{cpl}$ from Eq. \eqref{eq:vslipalt}. 

A last ingredient is necessary for the coupling model.
Given that the normal fluid is assumed incompressible  and that the hydrodynamic analogy of the 
GP equation gives a compressible fluid, we need to ensure the compatibility of the two flows and thus damp acoustic density  waves in the GP flow. A standard model  is  the so-called damped Gross-Pitaevskii equation \citep{QT-review-2017-tsubota-num} using a dissipation term controlled by a small dimensionless parameter $\eta_D$. We thus use  the following final modified GP equation:
\begin{eqnarray}
	\partial_t {\psi}  &=& i \left(\alpha \nabla^2 \psi -\gamma (|\psi|^2-{\overline{\rho}_s})\psi - \frac{1} {\alpha}\frac{({\bf v}_{\rm slip}^{cpl})^2}{4}\psi\right)\nonumber \\
	&-& ({\bf v}_{\rm slip}^{cpl}\cdot \nabla)\psi -\frac{1}{2} (\nabla\cdot {\bf v}_{\rm slip}^{cpl}) \psi \nonumber\\
	&+& \eta_D (\alpha \nabla^2 \psi -\gamma (|\psi|^2-{\overline{\rho}_s})\psi +\mu \psi).
	\label{Eq:GPEfullNum}
\end{eqnarray}
Parameters $\alpha=\frac{\hbar}{2 m}$ and $\gamma=\frac{g}{\hbar}$ are determined as usually from $c$ (see Eq. \eqref{Eq:defc}) and $\xi$ (see Eq. \eqref{Eq:defxi}), with $|\Psi|_0^2={\overline{\rho}_s}$.
Note that the value of $\alpha$ should be of order of the normal viscosity $\nu_n$. 
The initial-data wavefunction is normalized to $|\psi|^2={\overline{\rho}_s}$.
The term $\mu$ is introduced to ensure mass conservation in the modified GP equation.

\pagebreak

The final system of coupled equations \eqref{eq:NSEfullNum} and \eqref{Eq:GPEfullNum}  is advanced in time using a fourth-order Runge-Kutta method (with implicit discretization of Laplacian operators). Fourier-spectral  space discretization is used for both equations. The coupling algorithm was implemented in the framework of the modern parallel (MPI-OpenMP) numerical code called GPS (Gross-Pitaevskii Simulator) \citep{HPC-Parnaudeau-2015}. The GPS code was initially designed as a spectral parallel solver for the GP equation using various time-integration methods (Strang splitting, relaxation, Crank-Nicolson). It was recently used to simulate quantum turbulent flows \citep{KOBAYASHI2021}. The Navier-Stokes solver was added to the GPS code using standard Fourier pseudo-spectral method \citep{Got-Ors}.  Only one external library, FFTW \citep{HPC-FFTW}, was required for the computation. 

The coupling model has several coefficients that have to be fixed accordingly to the physics or be adjusted numerically. To give the model a physical background, the friction coefficients $U_\star$ and $V_\star$ were linked to tabulated experimental friction coefficients $B_{tab}$ and $B_{tab}'$ used in the physical literature for helium II. Equivalence relations  between friction coefficients are detailed in \ref{sec:Coef}. In the following, we prefer to set test cases using realistic values for $B_{tab}$ and $B_{tab}'$.  Normal $\rho_n$  and superfluid $\rho_s$ mass densities, the normal fluid viscosity $\nu_n$ are also fixed based on the physics of helium II, depending on the intermediate temperature between 0 and 2.17 K. 

The model also includes a few numerical coefficients  that have to be prescribed. These extra coefficients are the two smoothing parameters $\epsilon^2$ and $k_{\rm reg}$ used in the definition of ${\bf v}_s^{reg}$  \eqref{eq:vreg}, and the dissipation coefficient $\eta_D$ in \eqref{Eq:GPEfullNum}. On dimensional grounds, $\epsilon^2$ has to be proportional to $\overline{\rho}_S$, $k_{\rm reg}$ to $\xi^{-1}$ (the inverse of the healing length) and $\eta_D$ to the physical friction coefficient $B_{tab}$.
Remembering that $\overline{\rho}_S$ is close to the value $1$, it is consistent to use $\epsilon^2 = C_\epsilon \overline{\rho}_S$, with $C_\epsilon$ a small value constant. We set the second coefficient in a similar way, $k_{\rm reg} = C_k \xi^{-1}$, with $C_k \leq 1$ (as commonly set in simulations of GP quantum turbulence). The values of constants $C_\epsilon$ and $C_k$ will be adjusted in the next section by numerical tests reproducing 
the evolution of quantized vortices in a normal fluid.

\pagebreak

\section{Numerical results} \label{sec:Numr}

We use in this section classical cases of vortex dynamics to numerically validate the model. We adopt the following methodology:
\begin{itemize}
	\item The first preliminary test is more qualitative and intended to check that  the coupling model produces the correct displacement of a stationary 2D quantized vortex array. The vortex crystal  defined by \cite{Nore1994} is formed by two positive and two negative vortices in a 2D domain $[0, 2\pi]^2$, with center coordinates $(\frac{\pi}{2},\frac{\pi}{2})$ and $(\frac{3\pi}{2},\frac{3\pi}{2})$ for the positive ones,  and $(\frac{3\pi}{2},\frac{\pi}{2})$ et $(\frac{\pi}{2},\frac{3\pi}{2})$  for the negative vortices. This symmetric crystal arrangement has the property that ${\bf v}_s^{\rm reg}=0$, and consequently ${\bf v}_{L}= {\bf v}_{\rm slip}$ (see Fig. \ref{fig:sketch}).
	After obtaining the initial state of the crystal by solving the ARGLE equation \eqref{eq:ARGL} with ${\bf U}^{adv}=0$, it is possible to impose a constant velocity ${\bf v}_n$ to the normal fluid (i.e. the Navier-Stokes equations are not solved) and monitor how the crystal is deformed. For ${\bf v}_n$ directed following the $x$-axis the crystal remains stable and is translated by ${\bf v}_{\rm slip}$, while for  vertical ${\bf v}_n$, the crystal is deformed and a superfluid velocity ${\bf v}_s^{\rm reg}$ appears.  
	The obtained short term behavior (pictures not shown) of the vortex crystal corresponds to this expected motion and thus confirms  that the coupling model gives the correct  displacement of quantized vortices in an imposed constant normal flow. 
	
	\item The second numerical test is aimed at finely tune the parameters of the model ($\epsilon^2$,  $k_{\rm reg}$ and $\eta_D$) for a vortex configuration with non-trivial ${\bf v}_s^{\rm reg}$. For this purpose, we use the case of a 2D vortex dipole for which analytical solutions are available.  The one- or two-way GP-NS coupling could be tested using this benchmark. This case is described in detail in Sec. \ref{subsec:validation}.
	
	\item Once the values of the parameters are fixed, we test the complete coupling model by simulating the 3D dynamics of a superfluid vortex ring moving in a normal fluid. We then compare the results with those obtained by LV-NS coupling methods. We describe in Sec.  \ref{subsec:vortex-ring} the case of a single vortex ring and the case of the head-on collision of two vortex rings, moving in a normal fluid.

\end{itemize}

\subsection{2D superfluid vortex dipole and determination of model coefficients}
\label{subsec:validation}

We consider a superfluid vortex dipole in a periodic domain $[0,2\pi]^2$. The positive vortex (of circulation $\Gamma=\frac{h}{m} = 4\pi\alpha$) is initially centered at  $(x_+,y_+) = (x_0, \pi+\frac{R_0}{2})$ and the negative vortex (of circulation $-\Gamma$) at $(x_-,y_-) = (x_0, \pi-\frac{R_0}{2})$. The dipole is moving along the $x$-axis, symmetrically to the center line $y = \pi$. Parameters $x_0$ and $R_0$ define the initial streamwise position of the dipole and its initial radius, respectively. In absence of normal fluid, the  superfluid dipole translates in a periodic domain with known velocity \citep{Nazarenko2020}:
\begin{equation}\label{eq:R-us}
	{\bf u}_s = u_s {\bf e}_x, \quad u_s \approx  \frac{\Gamma}{4 \pi}\left(\frac{1+\cos(d)}{\sin(d)} + \frac{d}{\pi}\right),
\end{equation}
where $ d = 2R =y^+ - y^-$ is the distance between vortices. If a constant normal fluid velocity is imposed along the streamwise direction (${\bf u}_n = u_n {\bf e}_x$), the balance of forces acting on the dipole lead to the following analytical expressions for the horizontal $\dot{x}(t)$  and vertical $\dot{R}(t)$ velocities describing the dynamics of the dipole (see details in   \ref{sec:analytic 2D vortex dipole}):
\begin{align}
	\dot{x}(t) &= \frac{\gamma_0^2\rho_s\omega_s(u_n -u_s)}{(\gamma_0^2 +(\gamma_0' -\rho_s\omega_s)^2)(\rho_s\omega_s-\gamma_0')} 
	+\frac{u_s\rho_s\omega_s -\gamma_0'u_n}{\rho_s\omega_s-\gamma_0'},
	\label{eq:analytic_vel_x}\\
	\dot{R}(t) &= \frac{\gamma_0\rho_s\omega_s}{\gamma_0^2 +(\gamma_0' -\rho_s\omega_s)^2}(u_n -u_s),
	\label{eq:analytic_vel_R}
\end{align}
where $\gamma_0$, $\gamma'_0$ are two physical parameters related to the temperature and $\omega_s = (\nabla\times{\bf u}_s) {\bf e}_z$. Solution \eqref{eq:analytic_vel_x}-\eqref{eq:analytic_vel_R} is used in the following to finely tune the parameters of the coupling model.

\subsubsection{One-way GP-NS coupling}

\begin{figure}[!h]
	\begin{center}
		\centering
			\includegraphics[width=\textwidth]{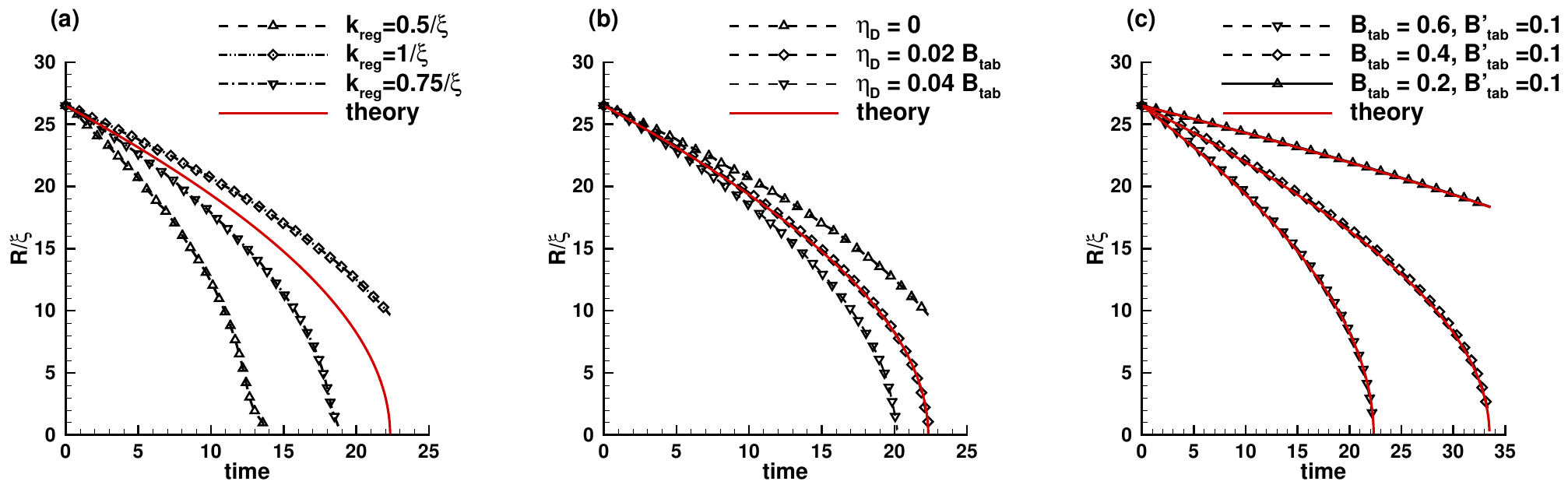}
		\caption{
			2D evolution of a superfluid vortex dipole. One-way GP-NS coupling, with ${\bf u}_n=0$. Time evolution of the half distance between the two vortices normalized by the size of the vortex core $\xi$. Solid lines represent the analytical solution. 
				(a) Results for three values of  of the smoothing wave number $k_{\rm reg}$ and common values $B_{tab}=0.6$ and $B'_{tab}=0.1$. (b) Results for three values  of the dissipation parameter $\eta_D$ and common values $B_{tab}=0.6$ and $B_{tab}=0.1$.
				(c) Results for $k_{\rm reg} =1/\xi$, $\eta_D =0.02B_{tab}$, and three different choices for the coupling force parameters $B_{tab}$ and $B'_{tab}$.
		}
		\label{fig:diffKernel_new}
	\end{center}
\end{figure}

We start by considering the one-way GP-NS coupling. The NS equations are not solved and we take $u_n=0$, which gives simpler relations for the analytical solution \eqref{eq:analytic_vel_x}-\eqref{eq:analytic_vel_R}. The superfluid vortex dipole is initially generated using the method  suggested by 
\cite{billam2014onsager} to impose the atomic density and the phase of the wave function.
 This case allows us to assess on the effect of the three parameters of the model:
\begin{itemize}
	
	\item The {\em regularization wave-number} $k_{\rm reg}$ is necessary in Eq. \eqref{eq:vreg} to obtain a smooth velocity ${\bf v}^{reg}_s$ and corresponding smooth vorticity ${\bf \Omega}$ in  Eq. \eqref{eq-Omega}.  It acts like a filter by smoothing the superfluid velocity and slightly diffusing the vorticity in the surrounding area of a vortex, which is the zone where the coupling force term is computed. 
	The choice of the regulation length scale $1/k_{\rm reg}$ was found to be critical to balance the accuracy and validity of the numerical simulation. If $k_{\rm reg}$ is too large (i. e. the vorticity around a vortex line is not smooth enough), the results could be more accurate, but the simulation might be unstable because of numerical oscillations (wiggles). On the other hand, if $k_{\rm reg}$ is too small, the numerical results are stable, but the accuracy is diminished. A trade-off between these two effects  thus should be found. Figure  \ref{fig:diffKernel_new} (a) shows that by decreasing $k_{\rm reg}$, the vortices of the dipole approach to each other with  an increasing rate. 
	We fixed $k_{\rm reg} = 1/\xi$, considering that a regularization length scale of the order the vortex core is physically reasonable.
	
	\item The {\em small parameter} $\epsilon^2$ in Eq. \eqref{eq:vreg} is also needed to avoid the singularity of the superfluid velocity  when the vortex line pass near a mesh node (as $\psib \psi$ is zero on the vortex line). We took $\epsilon^2 =0.1$ to ensure that the corresponding effective regularization length $0.31\xi$ is smaller than regularization length introduced by the $k_{\rm reg}$.
	
	\item The {\em dissipation} parameter $\eta_D$  was introduced in the GP equation to damp sound (pressure) waves and thus ensure the compatibility of the model with the incompressible flow assumption for both normal and superfluid. This dissipation effect is also affecting the intensity of the coupling force, which suggests that is reasonable to assume that $\eta_D$ is proportional to $B_{tab}$. Figure \ref{fig:diffKernel_new} (b) compares the numerical results with the analytical solution with different $\eta_D$. When setting $\eta_D=0$, we found  that the two vortices of the dipole  do not approach to each other fast enough and the gap between their positions do not evolve any more after reaching the value of approximately $10\xi$. When $\eta_D$ is increased,  vortices approach to each other in a increasing rate. The parameter $\eta_D$ was finally fixed to the value $0.02 B_{tab}$, for which the numerical solution fits perfectly to the analytical solution. When $B_{tab}=B'_{tab}$, the value $\eta_D=0.01 B_{tab}$ is also a good choice for the dissipation constant.
\end{itemize}

\begin{figure}[!ht]
	\begin{center}
		\centering
		\includegraphics[width=\textwidth]{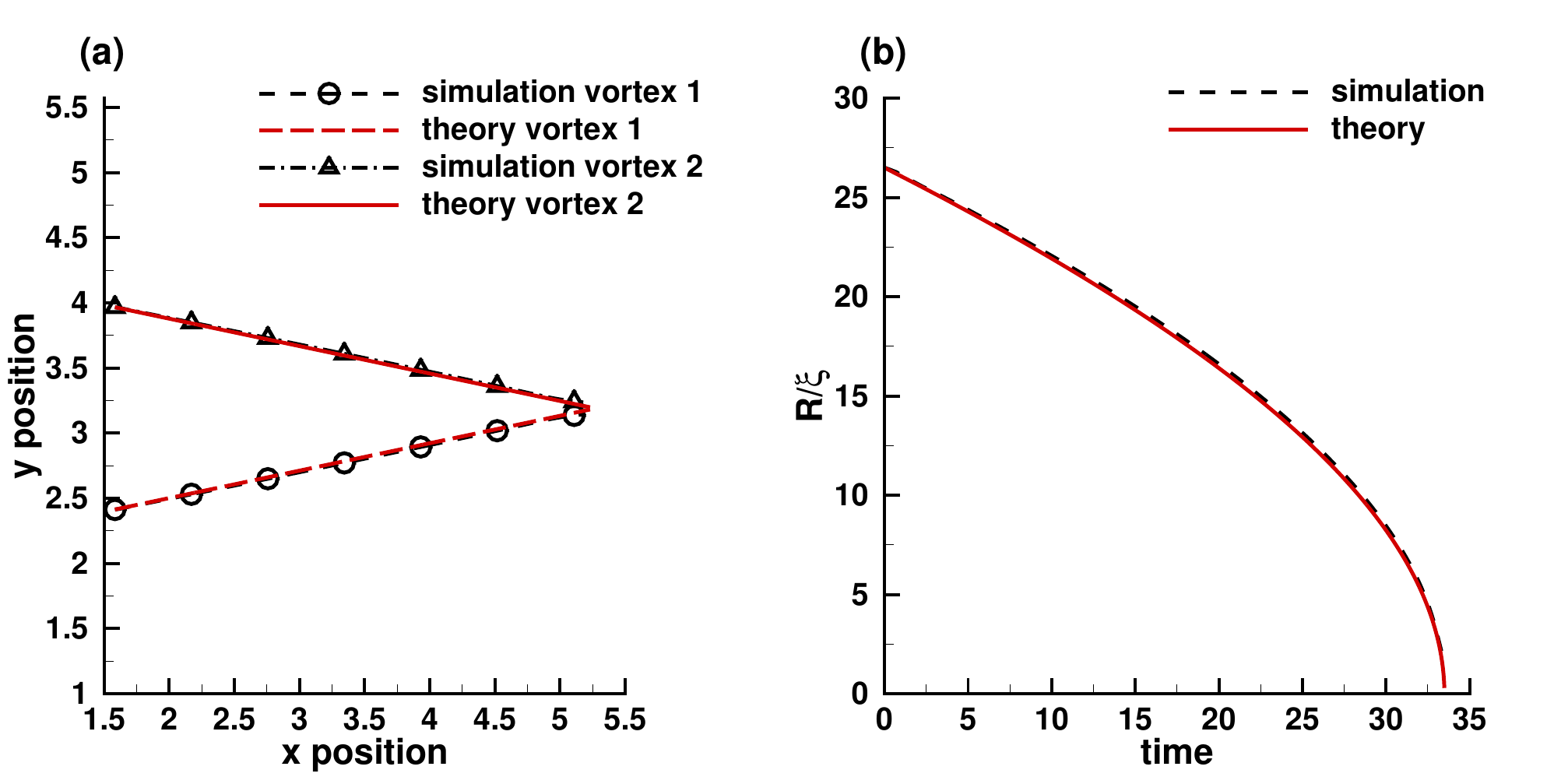}
		\caption{2D evolution of a superfluid vortex dipole. One-way GP-NS coupling, with ${\bf u}_n=0$.  Simulation with fixed parameters: $B_{tab}$ = 0.4, $B_{tab}' = 0.1$, $d/\xi = 53$, $N = 256$, $k_{\rm reg}^{-1}=\xi$, $\eta_D=0.02 B_{tab}$.
			(a) Trajectories of the two vortices. (b) Time evolution of the the half distance between the two vortices normalized by the size of the vortex core.}
		\label{fig:Vortex_pair_one_Bt_0_4_Bpt_0_1_1D}
	\end{center}
\end{figure}

Figure \ref{fig:diffKernel_new} (c) shows that using the values $\epsilon^2=0.1$,  $k_{\rm reg}=1/\xi$ and  $\eta_D=0.02B_{tab}$, the numerical results fit perfectly with the analytical solution for different coupling force coefficients $B_{tab}$ and $B'_{tab}$. Figure \ref{fig:Vortex_pair_one_Bt_0_4_Bpt_0_1_1D} offers a final validation of the values found for  the parameters of the model by depicting the time trajectories and time evolution of the radius of the dipole for typical values of coupling force coefficients $B_{tab}=0.4$ and $B'_{tab}=0.1$ (that will be used in the next sections).


\pagebreak

\subsubsection{Two-way GP-NS coupling}

\begin{figure}[!h]
	\begin{center}
		\centering
		
			\includegraphics[width=\textwidth]{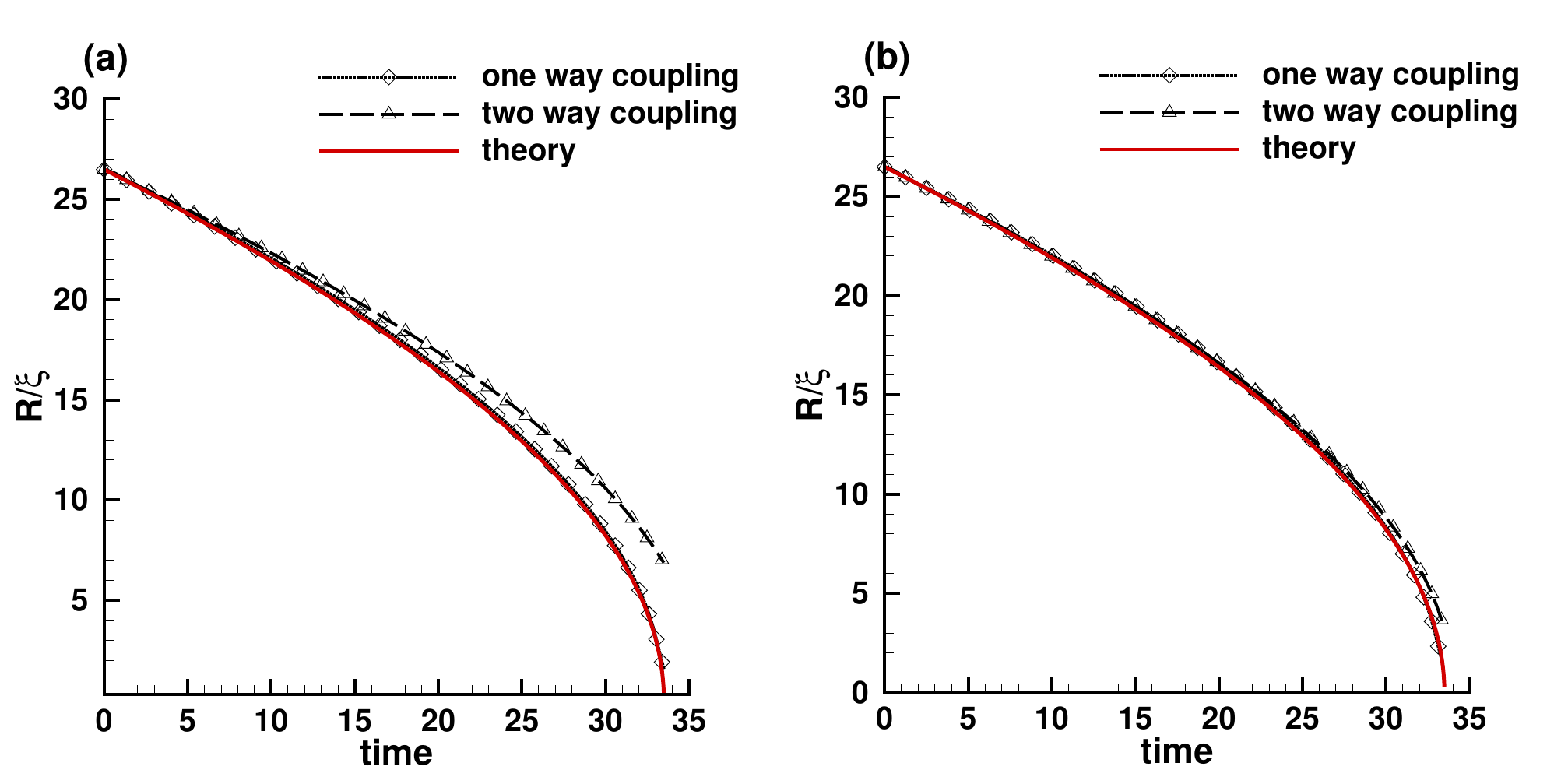}
		\caption{2D evolution of a superfluid vortex dipole.  Time evolution of the half distance between the two vortices normalized by the size of the vortex core $\xi$.  Comparison between ($-\diamond-$) one-way coupling (${\bf u}_n=0$) and ($-\triangle-$) two-way coupling (${\bf u}_n \neq 0$) for different physical parameters (a) : $B_{tab}$ = 0.4, $B'_{tab}$ = 0.1, $\eta_D=0.02 B_{tab}$, (b) : $B_{tab}$ = 0.4, $B'_{tab}$ = 0.4, $\eta_D=0.01 B_{tab}$.
			Common parameters of the model: $d/\xi = 53$, $N = 256$, $k_{\rm reg}^{-1}=\xi$.}
		\label{fig:Vortex_pair_comp_Np_256_1D}
	\end{center}
\end{figure}

\begin{figure}[!h]
	\begin{center}
		\centering
		
		\includegraphics[width=\textwidth]{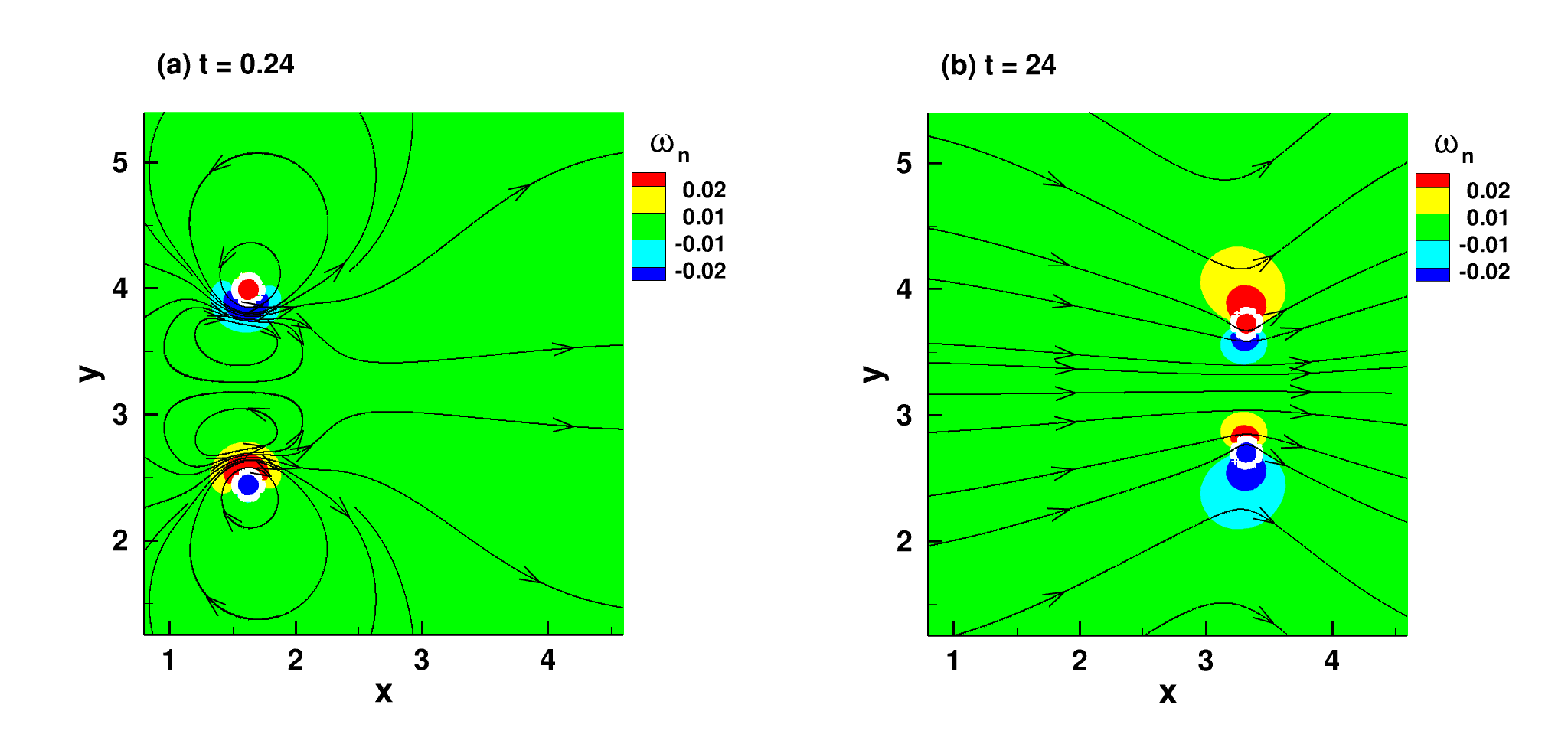}
		
		\caption{2D evolution of a superfluid vortex dipole. Two-way GP-NS coupling. Illustration of the triple-vortex structure of the flow. The entrained normal fluid is represented by its vorticity contours (colors) and streamlines (arrow black lines). Superfluid vortices (white circles) are identified by an iso-contour of low atomic density (0.5 $|\psi|^2_{max}$).  Snapshots of the flow for time instants: (a)  t=0.24, (b)  t=24. Parameters of the simulation: $B_{tab} = 0.4$, $B'_{tab} = 0.1$, $\eta_D=0$, $d/\xi = 53$, $N = 256$, $k_{\rm reg}^{-1}=\xi$.}
		\label{fig:2D_vort_triplering1}
	\end{center}
\end{figure}

We now simulate the time evolution of the same 2D dipole, but with the full two-way GP-NS coupling. The parameters of the model are kept the same as determined from the one-way coupling. The difference between the two types of coupling is visible in Fig. \ref{fig:Vortex_pair_comp_Np_256_1D}.  When considering the coupling force in  the NS equations (two-way coupling), the vortices of the dipole approach to each other with a reduced rate. This was expected, since the moving vortex dipole generates, through the coupling force, a normal fluid velocity ($u_n \neq 0$)  that finally  counteracts the mutual friction. The configuration of the flow is illustrated in Fig. \ref{fig:2D_vort_triplering1} presenting snapshots of the normal fluid vorticity and streamlines, together with the identification of the superfluid vortices by iso-contours of low-atomic density. We  observe a {\em triple-vortex-pair} structure consisting of 
a pair of superfluid anti-vortex and two pairs of anti-vortex of normal fluid: the first one is surrounding the superfluid vortices and rotates in the same direction, and the second one is adjacent to  superfluid vortices and rotates in the opposite direction. The stream lines show how  the normal fluid  is entrained by the motion of the superfluid vortex pair. By comparing the two snapshots, we can also observe that vortices move towards each other while translating downstream.

%

\pagebreak
\clearpage

\subsection{3D superfluid vortex ring}
\label{subsec:vortex-ring}

The triple-vortex structure observed in the 2D simulation of a vortex dipole is similar to that observed in the 3D flow generated by a superfluid vortex ring moving in a normal fluid \citep{kivotides2000triple,Giorgio2020,inui2021coupled}. We use this case to validate in 3D our full coupling GP-NS coupling model. The initial superfluid vortex ring is generated using Pad{\'e} approximations and the ARGLE procedure \citep{KOBAYASHI2021}. The normal fluid is initially at rest. 
%

\begin{figure}[!ht]
	\begin{center}
		\centering
			
				\includegraphics[width=0.47\textwidth]{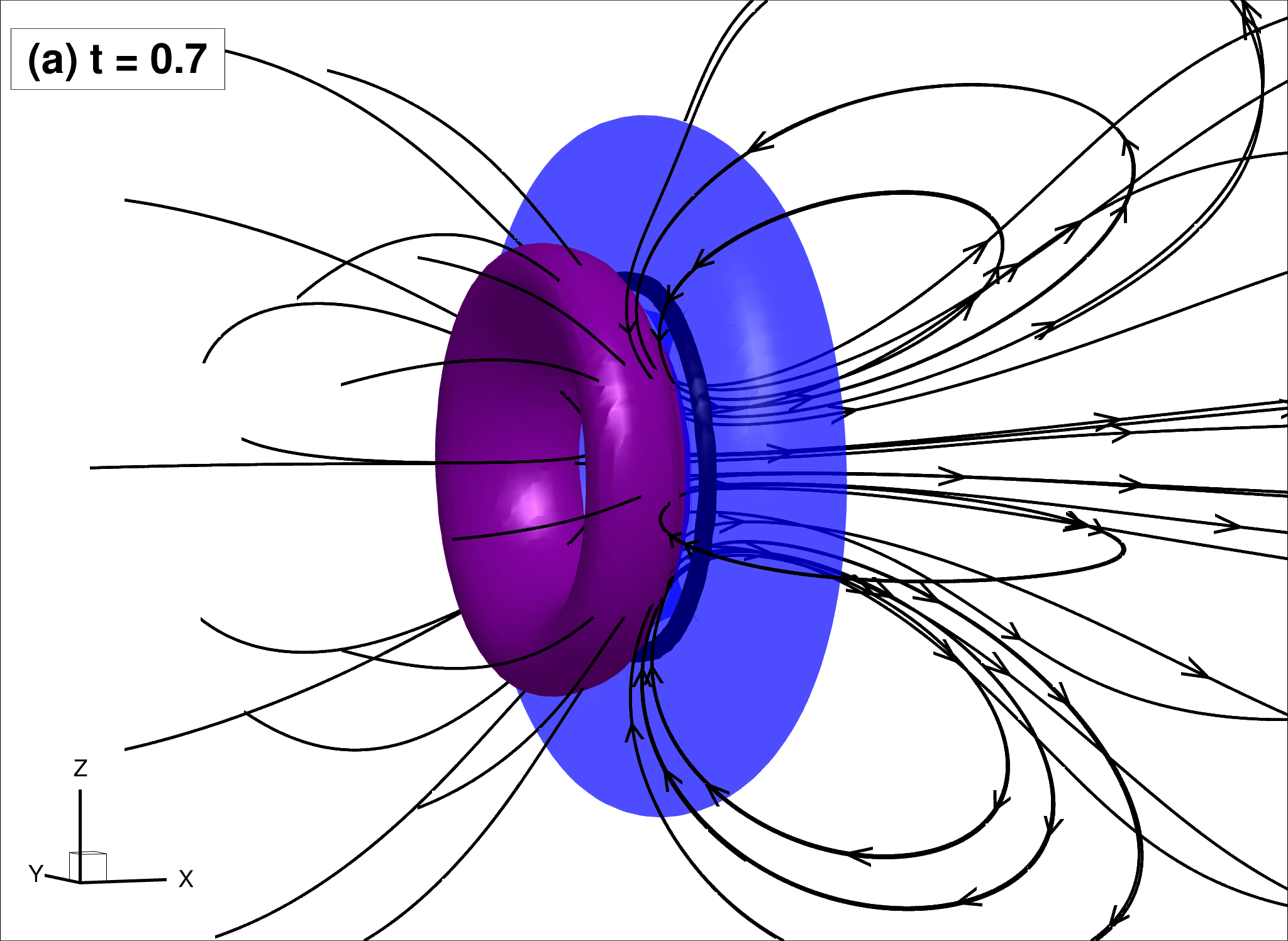}
			\includegraphics[width=0.47\textwidth]{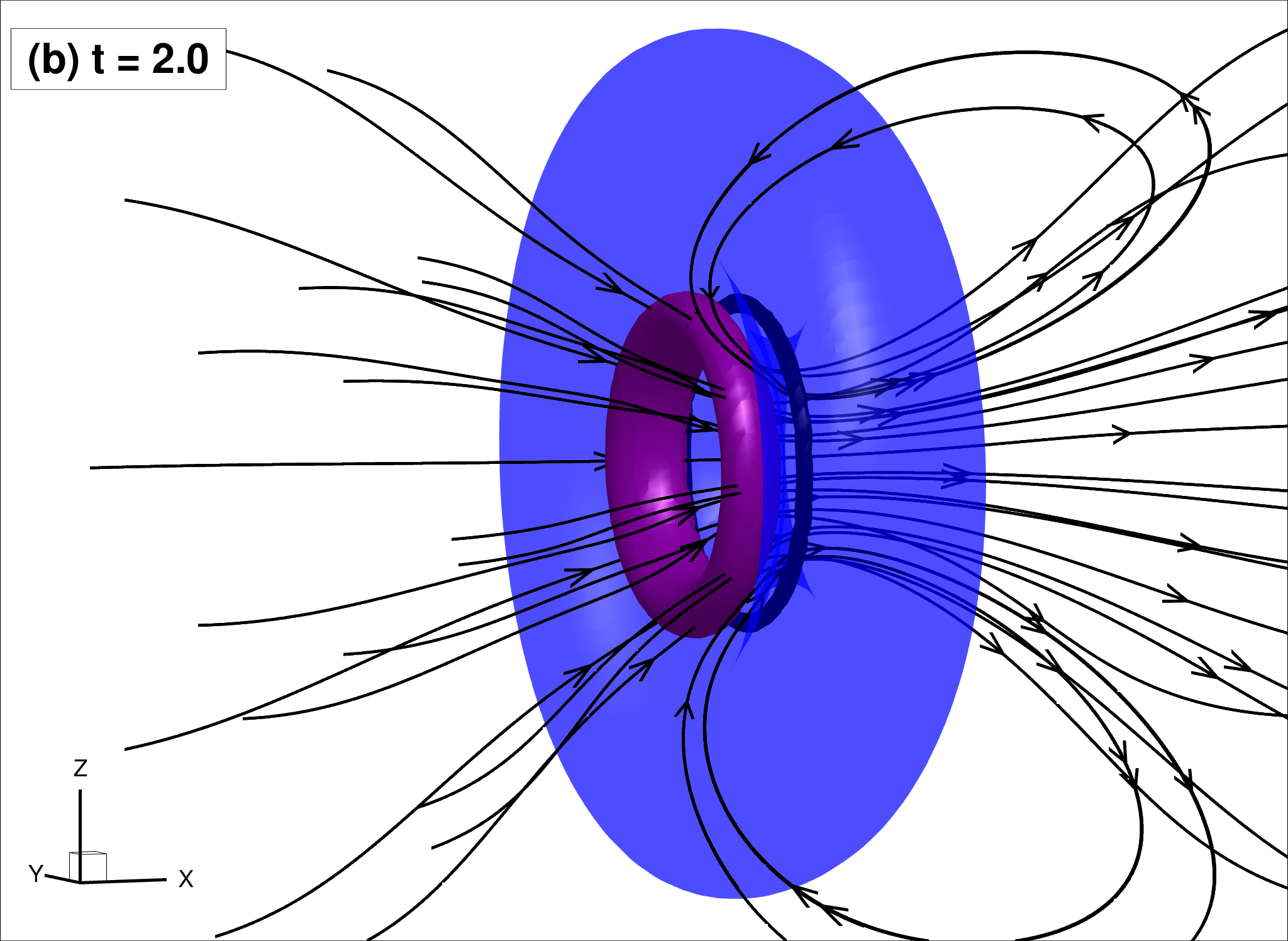}

				\includegraphics[width=0.47\textwidth]{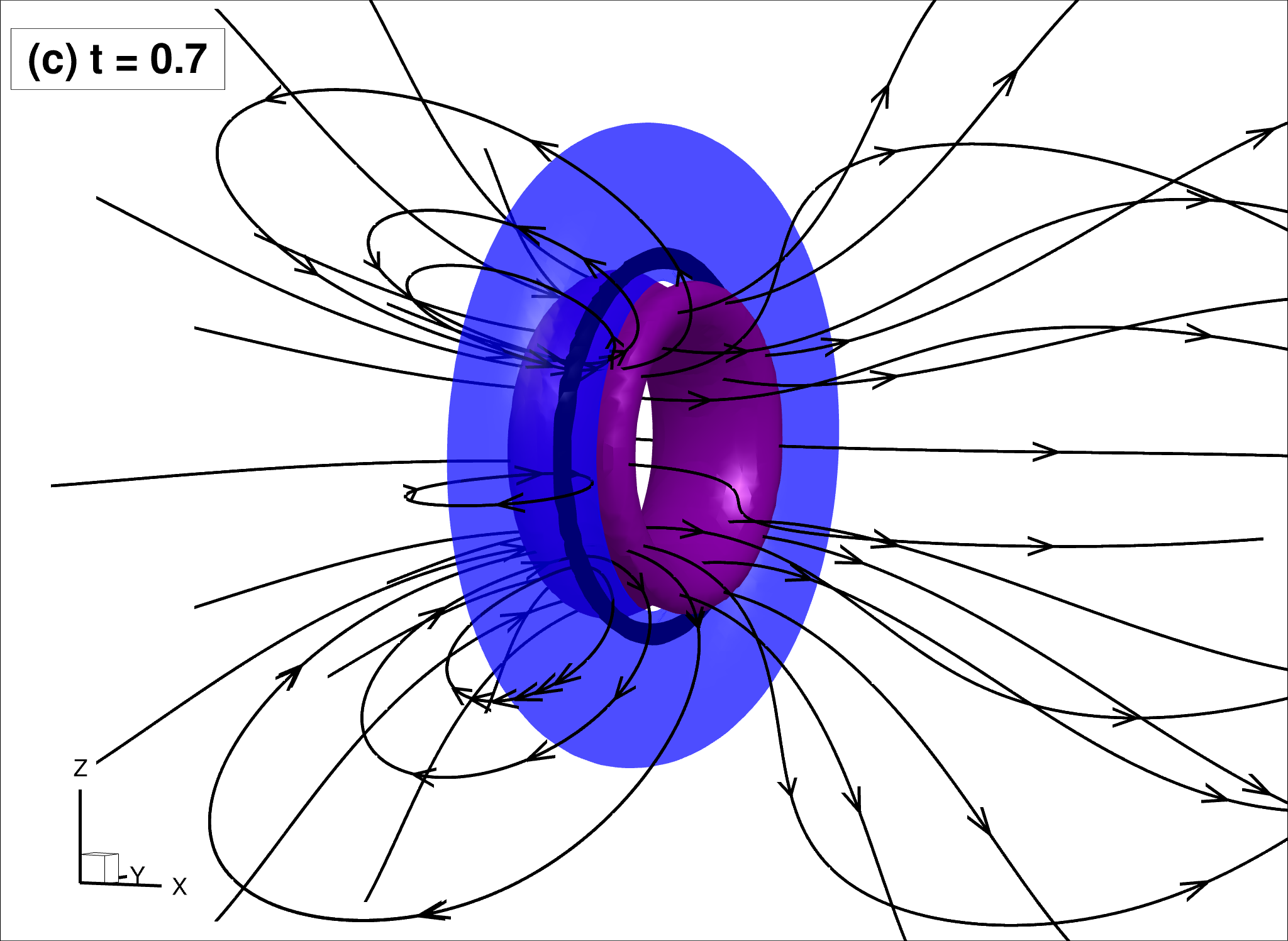}
\includegraphics[width=0.47\textwidth]{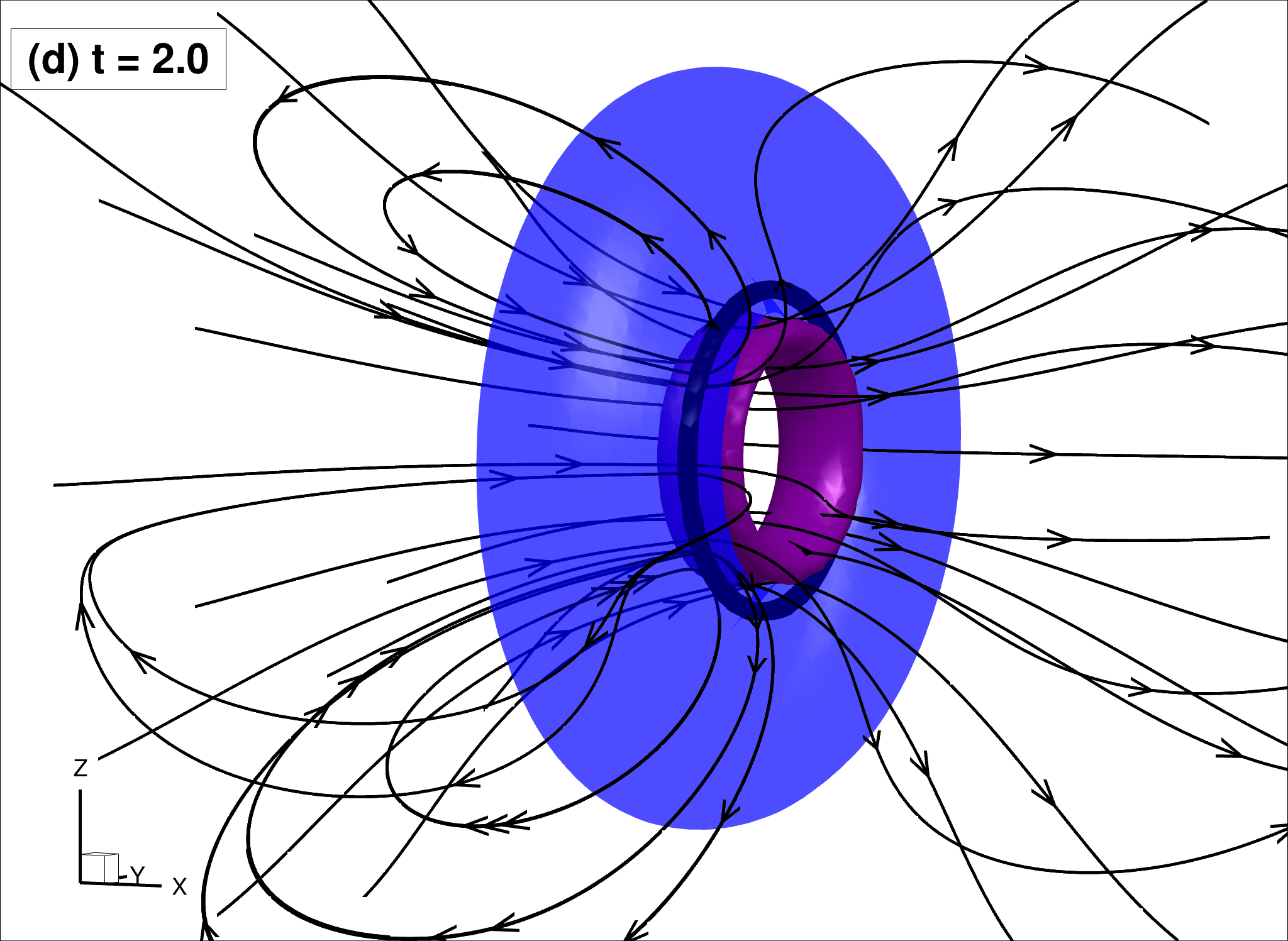}
			
		\caption{3D evolution of a superfluid vortex ring in a normal fluid initially at rest. Snapshots for two time instants. Physical parameters $\rho_n/\rho_s = 1$, $B_{tab}=B'_{tab} = 0.4$, $\eta_D=0.035B_{tab}$  (panels a, b),  $B_{tab}=0.4 > B'_{tab} = 0.1$, $\eta_D=0.05 B_{tab}$ (panels c, d). Illustration of the triple-vortex structure. The superfluid vortex ring  (in black) is identified by an iso-surface of low atomic density (0.5 $|\psi|^2_{max}$). The two counter-rotating normal vortex rings are identified by iso-surfaces of normal fluid azimuthal vorticity: $0.03$ for the blue outer ring and ($-0.03$) for the red inner ring. The streamlines in the normal fluid are also drawn. Mesh resolution $128^3$.}
		\label{fig:3D-triplerings-structure-Btab=Btabp}
	\end{center}
\end{figure}

Figure \ref{fig:3D-triplerings-structure-Btab=Btabp}  shows snapshots of the time evolution of the vortex ring for different physical parameters $B_{tab}=B'_{tab}$ (panels a, b) and $B_{tab} > B'_{tab}$ (panels c, d). 
The superfluid vortex ring (in black) moves in the $x$-direction from left to right and sweeps surrounding normal fluid due to the action of the coupling force. 
Two normal fluid vortex rings with opposite circulations are thus created, an outer one with large radius (in blue) and an inside smaller vortex ring (in red). The overall dimension of this triple-vortex rings structure reduces  while moving downstream. We thoroughly investigated the influence of the values of physical parameters on the topology of the triple-vortex. 
When $B_{tab} \approx B'_{tab}$ the small inner normal vortex ring (in red) travels at the rear of superfluid ring, while for $B_{tab} > B'_{tab}$ it is placed slightly in front of the superfluid ring.  

\begin{figure}[!ht]
	\begin{center}
		\centering
						\includegraphics[width=0.3\textwidth]{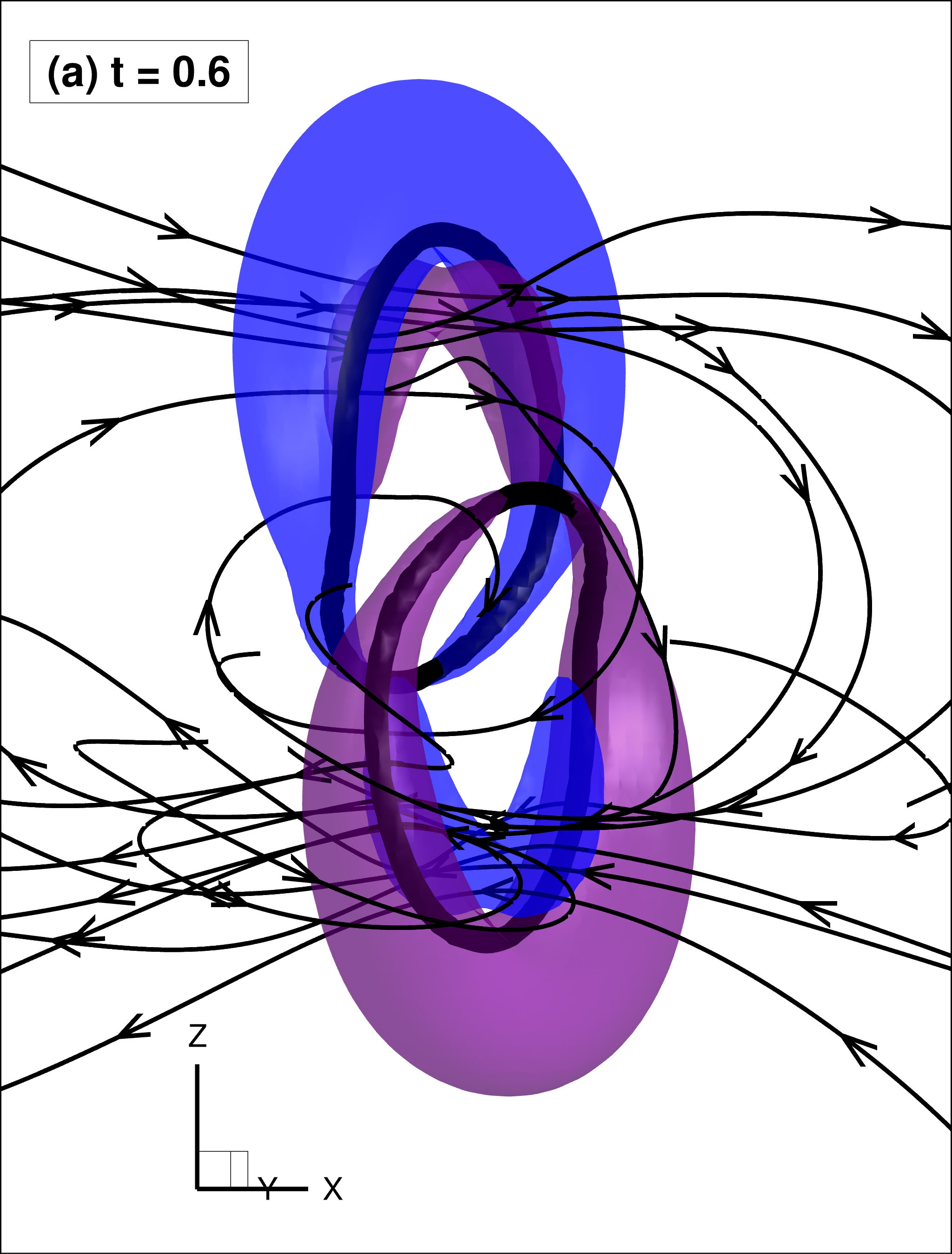}
								\includegraphics[width=0.3\textwidth]{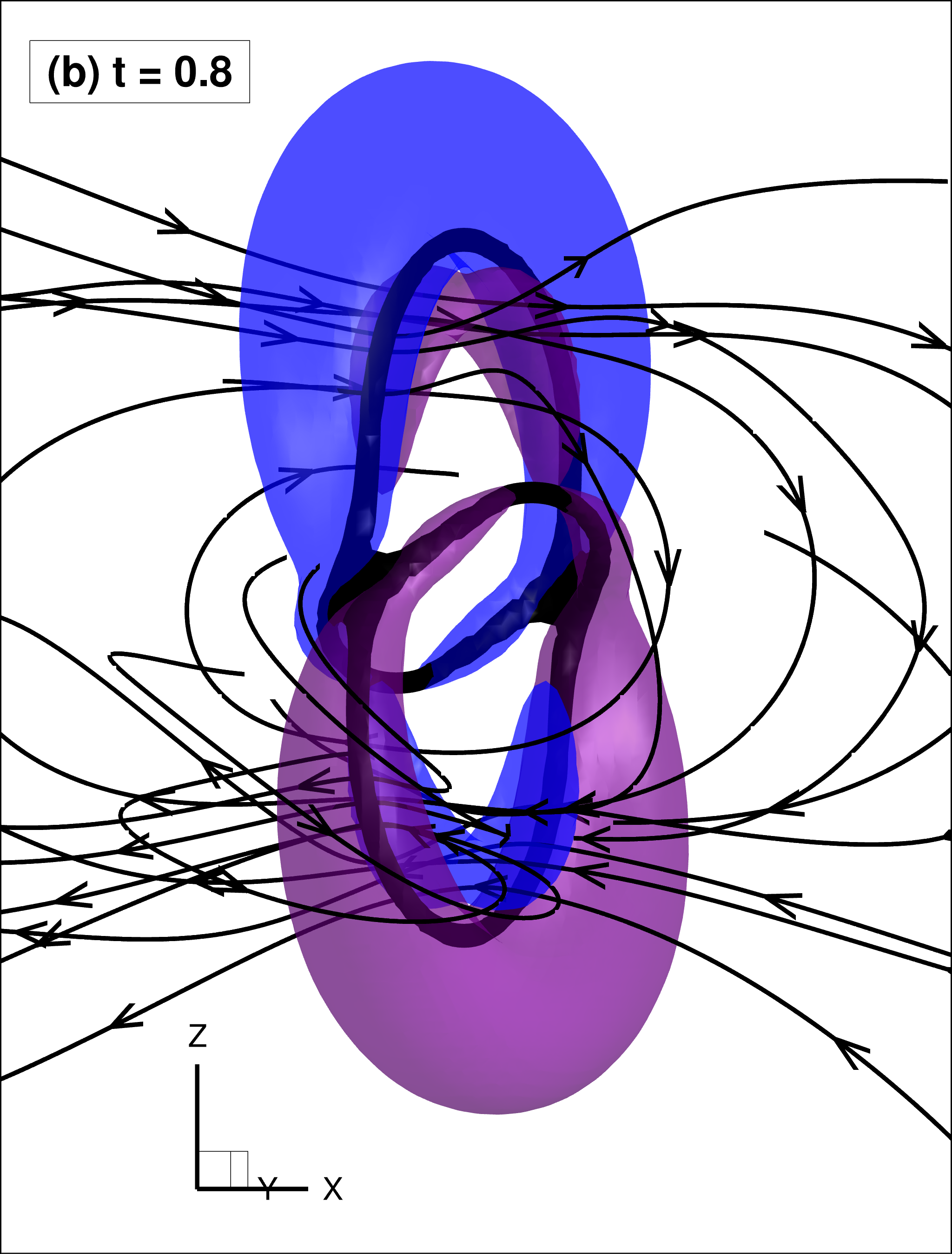}
										\includegraphics[width=0.3\textwidth]{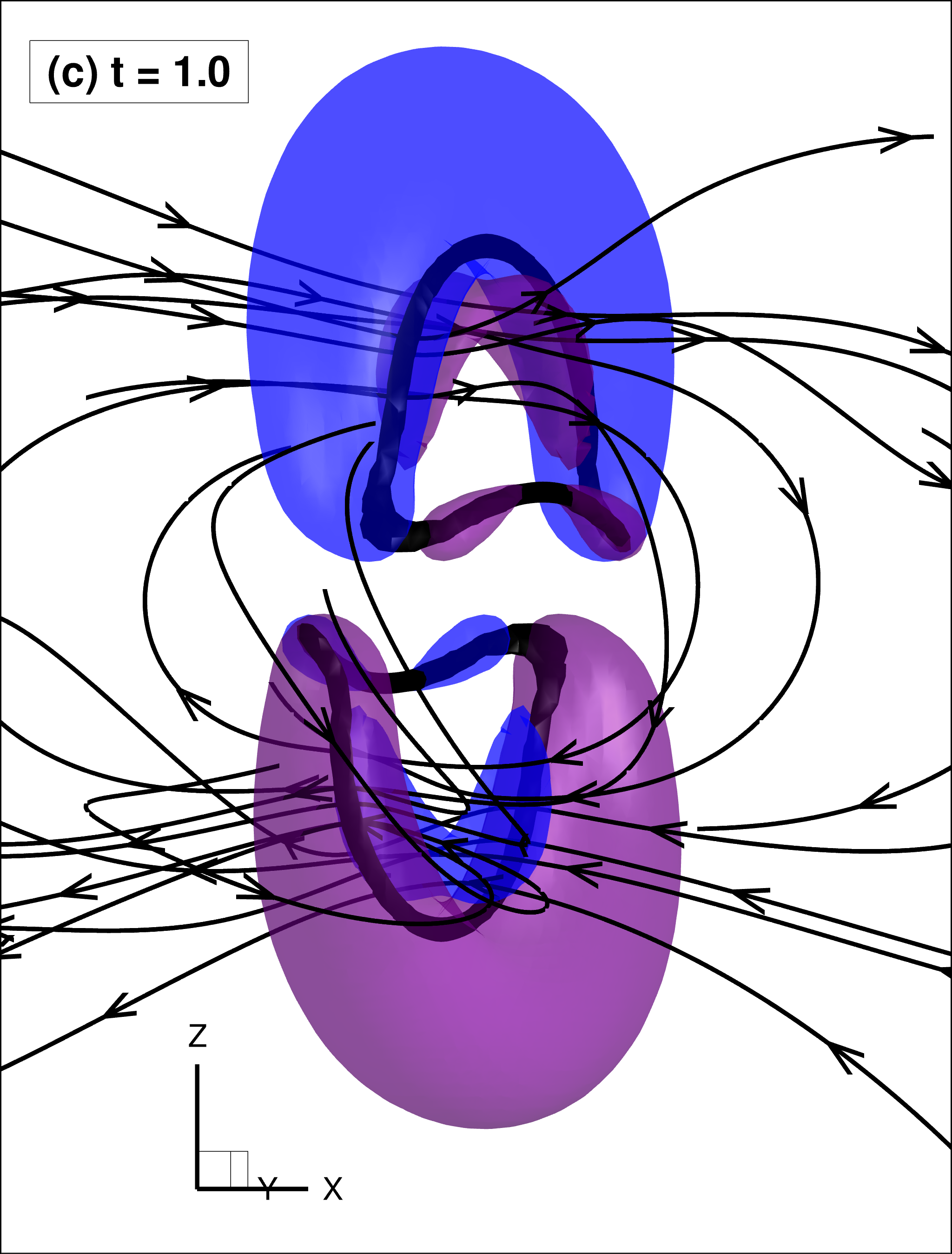}
		\caption{3D head-on collision of two superfluid vortex ring in a normal fluid initially at rest. Snapshots for three time instants.  Physical parameters $\rho_n/\rho_s = 1$, $B_{tab}=0.4$, $B'_{tab} = 0.1$, $\eta_D=0.05 B_{tab}$. Illustration of the structure of vortex reconnection.  The superfluid vortex ring  (in black) is identified by an iso-surface of low atomic density (0.2 $|\psi|^2_{max}$). The two counter-rotating normal vortex rings are identified by iso-surfaces of normal fluid azimuthal vorticity: $0.05$ for the blue outer ring and ($-0.05$) for the red inner ring. The streamlines in the normal fluid are also drawn. Mesh resolution $128^3$.}
		\label{fig:Intersection-3D-triplerings-structure-Btab>Btabp}
	\end{center}
\end{figure}

The triple-vortex ring structure illustrated in Fig.  \ref{fig:3D-triplerings-structure-Btab=Btabp}   is very similar to that recently found by LV-NS coupling models using vortex filaments for the superfluid and different NS solvers for the normal fluid \citep{Giorgio2020,inui2021coupled}. To emphasize the advantage of our GP-NS coupling to describe vortex interactions in superfluids without any phenomenological model, we also simulate the head-on collision of two superfluid vortex rings. In this case, superfluid vortex lines become distorted and their reconnection implies the exchange of parts of the lines and the formation of new tangled structures. This process is illustrated in Fig. \ref{fig:Intersection-3D-triplerings-structure-Btab>Btabp}. 
We use the same parameters as for the vortex ring case presented in Fig. \ref{fig:3D-triplerings-structure-Btab=Btabp} (c, d). Two vortex rings are seeded in the initial condition,  with the same radius and opposite propagation directions. Vortex centers are shifted along the vertical axis, as in the recent simulation by \cite{inui2021coupled}, using LV-NS coupling methods. The mutual induction deforms the vortex rings when they approach to each other (Fig.  \ref{fig:Intersection-3D-triplerings-structure-Btab>Btabp}a). The interaction  (Fig.  \ref{fig:Intersection-3D-triplerings-structure-Btab>Btabp}b) 
consists in the exchange of parts of each vortex line. After  reconnection (Fig.  \ref{fig:Intersection-3D-triplerings-structure-Btab>Btabp}c)  the two new vortex rings are distorted and continue their movement following their original direction. This complex interaction of superfluid vortex rings trigger in the normal fluid the formation of  two pairs of normal  vortex rings, that are attached to the  quantized vortex ring and undergo the well-known {\em cut-and-connect} reconnection mechanism for viscous NS vortex tubes \citep{fazle-1989,fazle-2011}. The obtained image of vortex interaction  is qualitatively similar to that obtained by  \cite{inui2021coupled} using phenomenological models for vortex reconnection, but there are differences. In particular, the repulsive motion observed  when the two vortex ring are getting closer and before the connection is more intense than in  LV-NS simulations. This affects the stretching of the normal fluid trapped between the two vortex rings. We recall that the superfluid vortex dynamics in our model obeys the GP equation,  without any phenomenological assumption on the reconnection process.

\pagebreak
\clearpage

\section{Conclusion} \label{sec:concl}

Recent models for the numerical simulation of two-fluid quantum flows (like helium II) were focused on coupling Navier-Stokes solvers for the normal fluid with vortex filaments methods for the superfluid fraction  \citep{Giorgio2020,inui2021coupled}. These models consider that the superfluid dynamics is essentially described by line-vortex interactions (Biot-Savart law) and thus referred to as LV-NS models. The resulting  main drawback is that vortex nucleation is absent from the description and superfluid vortex reconnections are necessarily based on phenomenological assumptions. We presented in this paper a model that links the Navier-Stokes (NS) normal flow dynamics to the Gross-Pitaevskii (GP) description of the superfluid fraction. The advantage of the model is that superfluid vortex dynamics is accurately and naturally described by the GP equation, as universally accepted in the literature.  The new GP-NS coupling model is compatible with physical concepts (mutual friction force, source term in NS) used in LV-NS models, but redefined in the framework of the GP superfluid dynamics. The modified GP equation follows some ideas introduced by \cite{Coste1998} to describe the compressible two-fluid liquid helium II, but it introduces new concepts: the regularized superfluid vorticity and velocity fields, the covariant gradient operator in the GP equation based on a slip velocity respecting the dynamics of vortex lines in the normal fluid. 

The new GP-NS coupling model was implemented in a  pseudo-spectral Fourier spectral code. Intensive tests validated the new numerical system against well-known benchmarks for the dynamics of different types or arrangements of quantized vortices (vortex crystal, vortex dipole and vortex rings) evolving in a normal fluid. The simulation of superfluid vortex head-on collision proved the ability of the method to account, without any phenomenological assumption, on the complex vortex interaction and reconnection. This new numerical model offers the possibility to revisit many fundamental phenomena established using the vortex filament method for superfluids (see the recent review by \cite{QT-review-2017-tsubota-num}): reconnections of superfluid vortex lines in a NS fluid, movement of superfluid vortex bundles in a normal fluid, counter-flow quantum turbulence and, finally, two-fluid quantum turbulence.

\section*{Acknowledgments}

The authors acknowledge financial support from the French ANR grant ANR-18-CE46-0013 QUTE-HPC. 
Part of this work was performed using computing resources of CRIANN 
(Centre R{\'e}gional Informatique et d'Applications Num{\'e}riques de Normandie, France).

\section*{References}


\begin{thebibliography}{32}
	\expandafter\ifx\csname natexlab\endcsname\relax\def\natexlab#1{#1}\fi
	\providecommand{\url}[1]{\texttt{#1}}
	\providecommand{\href}[2]{#2}
	\providecommand{\path}[1]{#1}
	\providecommand{\DOIprefix}{doi:}
	\providecommand{\ArXivprefix}{arXiv:}
	\providecommand{\URLprefix}{URL: }
	\providecommand{\Pubmedprefix}{pmid:}
	\providecommand{\doi}[1]{\href{http://dx.doi.org/#1}{\path{#1}}}
	\providecommand{\Pubmed}[1]{\href{pmid:#1}{\path{#1}}}
	\providecommand{\bibinfo}[2]{#2}
	\ifx\xfnm\relax \def\xfnm[#1]{\unskip,\space#1}\fi
	\bibitem[{Tisza(1938)}]{Tisza_1938}
	\bibinfo{author}{L.~Tisza},
	\newblock \bibinfo{title}{Transport phenomena in \uppercase{H}elium
		\uppercase{II}},
	\newblock \bibinfo{journal}{Nature} \bibinfo{volume}{141}
	(\bibinfo{year}{1938}) \bibinfo{pages}{913}.
	\bibitem[{Landau(1941)}]{Landau1941}
	\bibinfo{author}{L.~Landau},
	\newblock \bibinfo{title}{Theory of the superfluidity of \uppercase{H}elium
		\uppercase{II}},
	\newblock \bibinfo{journal}{Physical Review} \bibinfo{volume}{60}
	(\bibinfo{year}{1941}) \bibinfo{pages}{356--358}.
	\bibitem[{Nore et~al.(1997{\natexlab{a}})Nore, Abid, and Brachet}]{Nore97a}
	\bibinfo{author}{C.~Nore}, \bibinfo{author}{M.~Abid}, \bibinfo{author}{M.~E.
		Brachet},
	\newblock \bibinfo{title}{Decaying {K}olmogorov turbulence in a model of
		superflow},
	\newblock \bibinfo{journal}{Physics of Fluids} \bibinfo{volume}{9}
	(\bibinfo{year}{1997}{\natexlab{a}}) \bibinfo{pages}{2644--2669}.
	\bibitem[{Nore et~al.(1997{\natexlab{b}})Nore, Abid, and Brachet}]{Nore97b}
	\bibinfo{author}{C.~Nore}, \bibinfo{author}{M.~Abid},
	\bibinfo{author}{M.~Brachet},
	\newblock \bibinfo{title}{Kolmogorov turbulence in low-temperature superflows},
	\newblock \bibinfo{journal}{Phys. Rev. Letters} \bibinfo{volume}{78}
	(\bibinfo{year}{1997}{\natexlab{b}}) \bibinfo{pages}{3896--3899}.
	\bibitem[{Clark~di Leoni et~al.(2017)Clark~di Leoni, Mininni, and
		Brachet}]{Clark17}
	\bibinfo{author}{P.~Clark~di Leoni}, \bibinfo{author}{P.~D. Mininni},
	\bibinfo{author}{M.~E. Brachet},
	\newblock \bibinfo{title}{Dual cascade and dissipation mechanisms in helical
		quantum turbulence},
	\newblock \bibinfo{journal}{Phys. Rev. A} \bibinfo{volume}{95}
	(\bibinfo{year}{2017}) \bibinfo{pages}{053636}.
	\bibitem[{Kobayashi et~al.(2021)Kobayashi, Parnaudeau, Luddens, Lothod{\'e},
		Danaila, Brachet, and Danaila}]{KOBAYASHI2021}
	\bibinfo{author}{M.~Kobayashi}, \bibinfo{author}{P.~Parnaudeau},
	\bibinfo{author}{F.~Luddens}, \bibinfo{author}{C.~Lothod{\'e}},
	\bibinfo{author}{L.~Danaila}, \bibinfo{author}{M.~Brachet},
	\bibinfo{author}{I.~Danaila},
	\newblock \bibinfo{title}{Quantum turbulence simulations using the
		{G}ross-{P}itaevskii equation: High-performance computing and new numerical
		benchmarks},
	\newblock \bibinfo{journal}{Computer Physics Communications}
	\bibinfo{volume}{258} (\bibinfo{year}{2021}) \bibinfo{pages}{107579}.
	\bibitem[{Berloff et~al.(2014)Berloff, Brachet, and Proukakis}]{Berloff_2014}
	\bibinfo{author}{N.~G. Berloff}, \bibinfo{author}{M.~Brachet},
	\bibinfo{author}{N.~P. Proukakis},
	\newblock \bibinfo{title}{Modeling quantum fluid dynamics at nonzero
		temperatures},
	\newblock \bibinfo{journal}{Proceedings of the National Academy of Sciences}
	\bibinfo{volume}{111} (\bibinfo{year}{2014}) \bibinfo{pages}{4675--4682}.
	\bibitem[{Bekarevich and Khalatnikov(1961)}]{Bekarevich_Kalat_1961}
	\bibinfo{author}{I.~Bekarevich}, \bibinfo{author}{I.~Khalatnikov},
	\newblock \bibinfo{title}{Phenomenological derivation of the equations of
		vortex motion in \uppercase{H}elium \uppercase{II}},
	\newblock \bibinfo{journal}{Sov. Phys. JETP} \bibinfo{volume}{13}
	(\bibinfo{year}{1961}) \bibinfo{pages}{643}.
	\bibitem[{Hall et~al.(1956{\natexlab{a}})Hall, Vinen, and Shoenberg}]{HV1_1956}
	\bibinfo{author}{H.~E. Hall}, \bibinfo{author}{W.~F. Vinen},
	\bibinfo{author}{D.~Shoenberg},
	\newblock \bibinfo{title}{The rotation of liquid {H}elium {II.} {I.}
		{E}xperiments on the propagation of second sound in uniformly rotating
		{H}elium {II}},
	\newblock \bibinfo{journal}{Proceedings of the Royal Society of London. Series
		A. Mathematical and Physical Sciences} \bibinfo{volume}{238}
	(\bibinfo{year}{1956}{\natexlab{a}}) \bibinfo{pages}{204--214}.
	\bibitem[{Hall et~al.(1956{\natexlab{b}})Hall, Vinen, and Shoenberg}]{HV2_1956}
	\bibinfo{author}{H.~E. Hall}, \bibinfo{author}{W.~F. Vinen},
	\bibinfo{author}{D.~Shoenberg},
	\newblock \bibinfo{title}{The rotation of liquid {H}elium {II.} {II.} {T}he
		theory of mutual friction in uniformly rotating {H}elium {II}},
	\newblock \bibinfo{journal}{Proceedings of the Royal Society of London. Series
		A. Mathematical and Physical Sciences} \bibinfo{volume}{238}
	(\bibinfo{year}{1956}{\natexlab{b}}) \bibinfo{pages}{215--234}.
	\bibitem[{Adachi et~al.(2010)Adachi, Fujiyama, and Tsubota}]{Tsubota_2010}
	\bibinfo{author}{H.~Adachi}, \bibinfo{author}{S.~Fujiyama},
	\bibinfo{author}{M.~Tsubota},
	\newblock \bibinfo{title}{Steady-state counterflow quantum turbulence:
		Simulation of vortex filaments using the full {B}iot-{S}avart law},
	\newblock \bibinfo{journal}{Phys. Rev. B} \bibinfo{volume}{81}
	(\bibinfo{year}{2010}) \bibinfo{pages}{104511}.
	\bibitem[{Kivotides et~al.(2000)Kivotides, Barenghi, and
		Samuels}]{kivotides2000triple}
	\bibinfo{author}{D.~Kivotides}, \bibinfo{author}{C.~F. Barenghi},
	\bibinfo{author}{D.~C. Samuels},
	\newblock \bibinfo{title}{Triple vortex ring structure in superfluid helium
		{II}},
	\newblock \bibinfo{journal}{Science} \bibinfo{volume}{290}
	(\bibinfo{year}{2000}) \bibinfo{pages}{777--779}.
	\bibitem[{Galantucci et~al.(2020)Galantucci, Baggaley, Barenghi, and
		Krstulovic}]{Giorgio2020}
	\bibinfo{author}{L.~Galantucci}, \bibinfo{author}{A.~W. Baggaley},
	\bibinfo{author}{C.~F. Barenghi}, \bibinfo{author}{G.~Krstulovic},
	\newblock \bibinfo{title}{A new self-consistent approach of quantum turbulence
		in superfluid helium},
	\newblock \bibinfo{journal}{The European Physical Journal Plus}
	\bibinfo{volume}{135} (\bibinfo{year}{2020}) \bibinfo{pages}{547}.
	\bibitem[{Koplik and Levine(1993)}]{Koplik93}
	\bibinfo{author}{J.~Koplik}, \bibinfo{author}{H.~Levine},
	\newblock \bibinfo{title}{Vortex reconnection in superfluid helium},
	\newblock \bibinfo{journal}{Phys. Rev. Lett.} \bibinfo{volume}{71}
	(\bibinfo{year}{1993}) \bibinfo{pages}{1375--1378}.
	\bibitem[{Frisch et~al.(1992)Frisch, Pomeau, and Rica}]{FPR92}
	\bibinfo{author}{T.~Frisch}, \bibinfo{author}{Y.~Pomeau},
	\bibinfo{author}{S.~Rica},
	\newblock \bibinfo{title}{Transition to dissipation in a model of superflow},
	\newblock \bibinfo{journal}{Phys.Rev.Lett.} \bibinfo{volume}{69}
	(\bibinfo{year}{1992}) \bibinfo{pages}{1644}.
	\bibitem[{Balibar(2017)}]{Balibar2017}
	\bibinfo{author}{S.~Balibar},
	\newblock \bibinfo{title}{Laszlo {T}isza and the two-fluid model of
		superfluidity},
	\newblock \bibinfo{journal}{Comptes Rendus Physique} \bibinfo{volume}{18}
	(\bibinfo{year}{2017}) \bibinfo{pages}{586--591}. \bibinfo{note}{Science in
		the making: The Comptes rendus de l'Acad{\'e}mie des sciences throughout
		history}.
	\bibitem[{Coste(1998)}]{Coste1998}
	\bibinfo{author}{C.~Coste},
	\newblock \bibinfo{title}{Nonlinear {S}chr{\"o}dinger equation and superfluid
		hydrodynamics},
	\newblock \bibinfo{journal}{The European Physical Journal B - Condensed Matter
		and Complex Systems} \bibinfo{volume}{1} (\bibinfo{year}{1998})
	\bibinfo{pages}{245--253}.
	\bibitem[{Sonin(1997)}]{Sonin1997}
	\bibinfo{author}{E.~B. Sonin},
	\newblock \bibinfo{title}{Magnus force in superfluids and superconductors},
	\newblock \bibinfo{journal}{Phys. Rev. B} \bibinfo{volume}{55}
	(\bibinfo{year}{1997}) \bibinfo{pages}{485--501}.
	\bibitem[{Nore et~al.(1997)Nore, Abid, and Brachet}]{Nore1997}
	\bibinfo{author}{C.~Nore}, \bibinfo{author}{M.~Abid}, \bibinfo{author}{M.-E.
		Brachet},
	\newblock \bibinfo{title}{Decaying {K}olmogorov turbulence in a model of
		superflow},
	\newblock \bibinfo{journal}{Phys. Fluids} \bibinfo{volume}{9}
	(\bibinfo{year}{1997}) \bibinfo{pages}{2644--2669}.
	\bibitem[{Sandier and Serfaty(2007)}]{sandier-serfaty}
	\bibinfo{author}{E.~Sandier}, \bibinfo{author}{S.~Serfaty},
	\bibinfo{title}{Vortices in the Magnetic {G}inzburg-{L}andau Model},
	\bibinfo{publisher}{Birkhäuser Boston}, \bibinfo{year}{2007}.
	\bibitem[{Bao and Cai(2013)}]{BEC-review-2013-Bao-KRM}
	\bibinfo{author}{W.~Bao}, \bibinfo{author}{Y.~Cai},
	\newblock \bibinfo{title}{Mathematical theory and numerical methods for
		{B}ose-{E}instein condensation},
	\newblock \bibinfo{journal}{Kinetic and related models} \bibinfo{volume}{6}
	(\bibinfo{year}{2013}) \bibinfo{pages}{1--135}.
	\bibitem[{Tsubota et~al.(2017)Tsubota, Fujimoto, and
		Yui}]{QT-review-2017-tsubota-num}
	\bibinfo{author}{M.~Tsubota}, \bibinfo{author}{K.~Fujimoto},
	\bibinfo{author}{S.~Yui},
	\newblock \bibinfo{title}{Numerical studies of quantum turbulence},
	\newblock \bibinfo{journal}{J. of Low Temperature Physics}
	\bibinfo{volume}{188} (\bibinfo{year}{2017}) \bibinfo{pages}{119--189}.
	\bibitem[{Parnaudeau et~al.(2015)Parnaudeau, Suzuki, and
		Sac-Ep\'ee}]{HPC-Parnaudeau-2015}
	\bibinfo{author}{P.~Parnaudeau}, \bibinfo{author}{A.~Suzuki},
	\bibinfo{author}{J.-M. Sac-Ep\'ee},
	\newblock \bibinfo{title}{{GPS}: An efficient \& spectrally accurate code for
		computing {G}ross-{P}itaevskii equation},
	\newblock \bibinfo{journal}{ISC-2015, Research Posters Session}
	(\bibinfo{year}{2015}).
	\bibitem[{Gottlieb and Orszag(1977)}]{Got-Ors}
	\bibinfo{author}{D.~Gottlieb}, \bibinfo{author}{S.~A. Orszag},
	\bibinfo{title}{Numerical Analysis of Spectral Methods},
	\bibinfo{publisher}{SIAM}, \bibinfo{address}{Philadelphia},
	\bibinfo{year}{1977}.
	\bibitem[{Frigo and Johnson(2005)}]{HPC-FFTW}
	\bibinfo{author}{M.~Frigo}, \bibinfo{author}{S.~Johnson},
	\newblock \bibinfo{title}{Design and implementation of {FFTW3}},
	\newblock \bibinfo{journal}{Proceedings of the IEEE} \bibinfo{volume}{93}
	(\bibinfo{year}{2005}) \bibinfo{pages}{216--231}.
	\bibitem[{Nore et~al.(1994)Nore, Brachet, Cerda, and Tirapegui}]{Nore1994}
	\bibinfo{author}{C.~Nore}, \bibinfo{author}{M.~E. Brachet},
	\bibinfo{author}{E.~Cerda}, \bibinfo{author}{E.~Tirapegui},
	\newblock \bibinfo{title}{Scattering of first sound by superfluid vortices},
	\newblock \bibinfo{journal}{Phys. Rev. Lett.} \bibinfo{volume}{72}
	(\bibinfo{year}{1994}) \bibinfo{pages}{2593--2595}.
	\bibitem[{Griffin et~al.(2020)Griffin, Shukla, Brachet, and
		Nazarenko}]{Nazarenko2020}
	\bibinfo{author}{A.~Griffin}, \bibinfo{author}{V.~Shukla},
	\bibinfo{author}{M.-E. Brachet}, \bibinfo{author}{S.~Nazarenko},
	\newblock \bibinfo{title}{Magnus-force model for active particles trapped on
		superfluid vortices},
	\newblock \bibinfo{journal}{Phys. Rev. A} \bibinfo{volume}{101}
	(\bibinfo{year}{2020}) \bibinfo{pages}{053601}.
	\bibitem[{Billam et~al.(2014)Billam, Reeves, Anderson, and
		Bradley}]{billam2014onsager}
	\bibinfo{author}{T.~P. Billam}, \bibinfo{author}{M.~T. Reeves},
	\bibinfo{author}{B.~P. Anderson}, \bibinfo{author}{A.~S. Bradley},
	\newblock \bibinfo{title}{{O}nsager-{K}raichnan condensation in decaying
		two-dimensional quantum turbulence},
	\newblock \bibinfo{journal}{Phys. Rev. Lett.} \bibinfo{volume}{112}
	(\bibinfo{year}{2014}) \bibinfo{pages}{145301}.
	\bibitem[{Inui and Tsubota(2021)}]{inui2021coupled}
	\bibinfo{author}{S.~Inui}, \bibinfo{author}{M.~Tsubota},
	\newblock \bibinfo{title}{Coupled dynamics of quantized vortices and normal
		fluid in superfluid $^{4}\mathrm{He}$ based on the lattice {B}oltzmann
		method},
	\newblock \bibinfo{journal}{Phys. Rev. B} \bibinfo{volume}{104}
	(\bibinfo{year}{2021}) \bibinfo{pages}{214503}.
	\bibitem[{Melander and Hussain(1989)}]{fazle-1989}
	\bibinfo{author}{M.~V. Melander}, \bibinfo{author}{F.~Hussain},
	\newblock \bibinfo{title}{Cross‐linking of two antiparallel vortex tubes},
	\newblock \bibinfo{journal}{Physics of Fluids A: Fluid Dynamics}
	\bibinfo{volume}{1} (\bibinfo{year}{1989}) \bibinfo{pages}{633--636}.
	\bibitem[{Hussain and Duraisamy(2011)}]{fazle-2011}
	\bibinfo{author}{F.~Hussain}, \bibinfo{author}{K.~Duraisamy},
	\newblock \bibinfo{title}{Mechanics of viscous vortex reconnection},
	\newblock \bibinfo{journal}{Physics of Fluids} \bibinfo{volume}{23}
	(\bibinfo{year}{2011}) \bibinfo{pages}{021701}.
	\bibitem[{Barenghi et~al.(1983)Barenghi, Donnelly, and Vinen}]{Barenghi1983}
	\bibinfo{author}{C.~F. Barenghi}, \bibinfo{author}{R.~J. Donnelly},
	\bibinfo{author}{W.~F. Vinen},
	\newblock \bibinfo{title}{Friction on quantized vortices in {H}elium {II}. {A}
		review},
	\newblock \bibinfo{journal}{Journal of Low Temperature Physics}
	\bibinfo{volume}{52} (\bibinfo{year}{1983}) \bibinfo{pages}{189--247}.
	
\end{thebibliography}

\pagebreak

\appendix

\small
	\section{Expression of the slip velocity}\label{sec:app_formulae}
	To solve Eq. \eqref{eq:vslip5}, we use that  ${\bf v}_{\rm slip}$ is perpendicular to the vortex line:  $ {\bf v}_{\rm slip} \cdot {\bf \Omega}=0$. We obtain:
	\begin{equation}
		-\rho_s {\bf v}_{\rm slip} + \rho_n   ( B_\star  ( F{\bf \Omega}\times ( {\bf w}_p - {\bf v}_{\rm slip}))+B'_\star ({\bf w}_p - {\bf v}_{\rm slip})) =0,\label{eq:vslip6}
	\end{equation}
	or
	\begin{equation}
		- (\rho_s{+} B'_\star  \rho_n)   {\bf v}_{\rm slip}- \rho_n   B_\star F{\bf \Omega} \times {\bf v}_{\rm slip}
		=- \rho_n   ( B_\star F{\bf \Omega} \times {\bf w} + B'_\star {\bf w}_p ).
	\end{equation}
	Setting
	\begin{equation}
		{\bf v}_{\rm slip}=U_\star {\bf w}_p +V_\star F {\bf \Omega}\times{\bf w},
	\end{equation}
	we  obtain that
	\begin{eqnarray}
		&{-}&(\rho_s{+} B'_\star  \rho_n) (U_\star {\bf w}_p +V_\star F {\bf \Omega}\times{\bf w})\nonumber\\
		&-& \rho_n   B_\star F{\bf \Omega} \times (U_\star {\bf w}_p +V_\star F {\bf \Omega}\times{\bf w})\nonumber\\
		&=&- \rho_n   ( B_\star F {\bf \Omega} \times {\bf w} + B'_\star {\bf w}_p )
	\end{eqnarray}
	or, using that ${\bf \Omega} \times ({\bf \Omega} \times {\bf w})= {\bf \Omega} \times ({\bf \Omega} \times {\bf w}_p)=-{\bf \Omega}\cdot {\bf \Omega} \; {\bf w}_p$,
	\begin{eqnarray}
		&{-}&(\rho_s{+} B'_\star \rho_n) (U_\star {\bf w}_p +V_\star F {\bf \Omega}\times{\bf w})\nonumber\\
		&-& \rho_n   B_\star F U_\star {\bf \Omega} \times {\bf w}_p + \rho_n   B_\star F V_\star F {\bf \Omega}\cdot {\bf \Omega} \; {\bf w}_p\nonumber\\
		&=&
		- \rho_n   ( B_\star F{\bf \Omega} \times {\bf w} + B'_\star {\bf w}_p ).
	\end{eqnarray}
	Taking the inner product with  ${\bf w}_p$ and ${\bf \Omega}\times{\bf w}$, we infer that 
	\begin{eqnarray}
		{-}(\rho_s{+} B'_\star \rho_n) U_\star  + \rho_n   B_\star  V_\star F^2 {\bf \Omega}\cdot {\bf \Omega} \; &=&
		- \rho_n   B'_\star,   
		\\ 
		- \rho_n   B_\star U_\star {-}(\rho_s{+} B'_\star  \rho_n) V_\star &=&
		- \rho_n   B_\star. 
	\end{eqnarray}
	The final solution is 
	\begin{eqnarray}
		U_\star &=&\frac{\rho_n \left(B_\star^2 F^2  {\bf \Omega}\cdot {\bf \Omega}\rho_n {+}B'_\star \left(\rho_s{+}
			\rho_n B'_\star \right)\right)}{B_\star^2 F^2  {\bf \Omega}\cdot {\bf \Omega}\rho_n^2+\left(\rho_s{+}
			\rho_n B'_\star \right)^2},\\ 
		V_\star &=&\frac{{+}B_\star \rho_n \rho_s}{B_\star^2 F^2
			{\bf \Omega}\cdot {\bf \Omega}\rho_n^2+\left(\rho_s{+} \rho_n B'_\star \right)^2}.
	\end{eqnarray}

	\section{Expressions of friction coefficients}
	\label{sec:Coef}
	
	Friction coefficients $U_\star$ and $V_\star$ in Eq. \eqref{eq-vslip-final} can be related to physical friction coefficients tabulated for superfluid helium II \citep{Barenghi1983}. We recall that  three different scales appear in our model: the healing length $\xi$ that is also the scale of the vortex core, the smallest normal fluid length (the phonon/rotons mean free path) $\lambda$ and the inter-vortex distance $\ell$. In LV-NS models, since a vortex is a filament, $\xi=0$ and $\lambda<\ell$.
	In HBVK models, the superfluid vorticity  is averaged over a length scale larger than $\ell$.
	In our description, we average over some intermediate scale $l$, supposing that $\xi<\lambda<l<\ell$, to obtain a  mesoscopic description of the mutual friction.
	
	We consider the case of a uniform rotation and use the experimentally tabulated coefficients $B_{tab}$ and $B'_{tab}$ given by \cite{Barenghi1983}.
	By averaging over the vortex array, we must identify:
	\begin{eqnarray}
		B_{tab}&=& \frac{\rho} {\rho_n} V_\star,\\
		B'_{tab}&=& \frac{\rho} {\rho_n} U_\star. 
	\end{eqnarray}
	The above relations for ($U_\star$,  $V_\star)$ can be inverted into
	\begin{eqnarray}
		B_\star&=&\frac{\rho_s V_\star}{\rho_n \left( {\bf \hat \Omega}^2 V_\star^2+(U_\star-1)^2\right)},\\ 
		B'_\star&=&-\frac{\rho_s \left( {\bf \hat \Omega}^2 V_\star^2+(U_\star-1) U_\star\right)}{\rho_n
			\left( {\bf \hat \Omega}^2 V_\star^2+(U_\star-1)^2\right)}.
	\end{eqnarray}
	Using the that ${\bf \hat \Omega}^2=1$ on vortex lines,  we finally find for our coefficients $B_\star$,  $B'_\star$:
	\begin{eqnarray}
		B_\star&=&\frac{\rho_s \frac{\rho_n} {\rho} B_{tab}}{\rho_n \left( ( \frac{\rho_n} {\rho} B_{tab})^2+(\frac{\rho_n} {\rho} B'_{tab}-1)^2\right)},\\ 
		B'_\star&=&-\frac{\rho_s \left(  (\frac{\rho_n} {\rho} B_{tab})^2+(\frac{\rho_n} {\rho} B'_{tab}-1) \frac{\rho_n} {\rho} B'_{tab}\right)}{\rho_n
			\left(  (\frac{\rho_n} {\rho} B_{tab})^2+(\frac{\rho_n} {\rho} B'_{tab}-1)^2\right)},
	\end{eqnarray}
	or
	\begin{eqnarray}
		B_\star&=&\frac{ B_{tab}  \rho  \rho_s}{\rho_n^2
			\left( {B_{tab} '}^2+ B_{tab} ^2\right)-2  {B_{tab} '} \rho  \rho_n+\rho ^2},\\ 
		B'_\star&=&\frac{ {B_{tab} '} \rho  \rho_s-\rho_n \rho_s
			\left( {B_{tab} '}^2+ B_{tab} ^2\right)}{\rho_n^2
			\left( {B_{tab} '}^2+ B_{tab} ^2\right)-2  {B_{tab} '} \rho  \rho_n+\rho ^2}.
	\end{eqnarray}
	\section{Movement of a 2D vortex dipole in a normal fluid}
	\label{sec:analytic 2D vortex dipole}

We consider the superfluid vortex dipole described in Sec. \ref{subsec:validation}.
In a periodic domain and in absence of normal fluid, the  superfluid dipole moves in the $x$-direction with the velocity given by Eq. \eqref{eq:R-us} \citep{Nazarenko2020}. Considering that the vortices of the dipole are straight lines perpendicular to the movement plane $(x, y)$, we can apply the force balance equation \eqref{eq:fsnmag}. We assume that the velocity induced by the vortex line is (in the vicinity of the line):
\begin{equation}
	{\bf v}_s = u_s {\bf e}_x,
\end{equation}
with $u_s$ given by Eq. \eqref{eq:R-us}. Since the velocity of vortex lines is:
\begin{equation}
	{\bf v}_L = \dot{x}\, {\bf e}_x +  \dot{R}\, {\bf e}_y,
\end{equation}
the tangent vector is $s'={\bf e}_z$ and the vorticity  ${\bf \Omega} ={\omega}_s {\bf e}_z$, we can use Eqs.  \eqref{eq:mag}, \eqref{eq:fsnmag}, and  \eqref{eq:FSN}  to obtain:
\begin{equation}
	0 =  \rho_s \omega_s {\bf e}_z \times({\bf v}_L - {\bf v}_s) + \gamma_0 ({\bf v}_n-{\bf v}_L) +\gamma_0' {\bf e}_z \times({\bf v}_n - {\bf v}_L),
	\label{eq:balance force2D}
\end{equation}
with $\gamma_0 =  \rho_n B_\star/F$ and $\gamma_0' = - \rho_n B'_\star $. Assuming that ${\bf v}_n=u_n {\bf e}_x$, we separate from relation
 \eqref{eq:balance force2D} the two linear equations corresponding to $x$ and $y$ directions, respectively: 
\begin{align}
	0 &= -\rho_s\dot{R}\omega_s + \gamma_0(u_n-\dot{x}) + \dot{R}\gamma_0',\\
	0 &= \rho_s\omega_s(\dot{x}-u_s) -\gamma_0\dot{R} + \gamma_0'(u_n-\dot{x}).
\end{align}
The solution is obtained in the form:
\begin{align}
	\dot{x}(t)& = \frac{\gamma_0^2\rho_s\omega_s(u_n -u_s)}{(\gamma_0^2 +(\gamma_0' -\rho_s\omega_s)^2)(\rho_s\omega_s-\gamma_0')}
	+\frac{u_s\rho_s\omega_s -\gamma_0'u_n}{\rho_s\omega_s-\gamma_0'},\\
		\dot{R}(t) &= \frac{\gamma_0\rho_s\omega_s}{\gamma_0^2 +(\gamma_0' -\rho_s\omega_s)^2}(u_n -u_s).
\end{align}
To follow the position $x(t)$ and radius $R(t)$ of the dipole in time, we calculate:
\begin{equation}\label{eq:Rp}
		x(t) =\int_{0}^{t} \dot{x}(s)ds, \quad R(t) =\int_{0}^{t} \dot{R}(s)ds.
\end{equation}

\end{document}